\documentclass[12pt,a4paper]{article}

\usepackage{graphicx}
\def\bc{\begin{center}}
\def\ec{\end{center}}
\def\beq{\begin{equation}}
\def\eeq{\end{equation}}
\def\beq*{\begin{equation*}}
\def\eeq*{\end{equation*}}

\def\hs#1{\hspace*{#1cm}}
\def\av#1{\langle #1 \rangle}
\def\avr#1#2{\langle {#1} \rangle^{}_{#2}}
\def\omN{{\omega_N}}

\def\ommu{{\omega_\mu}}
\def\omn{{\omega_n}}

\def\muF{{\mu_F^{}}}
\def\nF{{n_F^{}}}

\def\nB{{n_B^{}}}

\def\cov{{\textrm{cov}}}

\def\bc{\begin{center}}
\def\ec{\end{center}}
\def\beq{\begin{equation}}
\def\eeq{\end{equation}}

\def\hs#1{\hspace*{#1cm}}
\def\av#1{\langle #1 \rangle}

\def\avr#1#2{\langle {#1} \rangle^{}_{#2}}

\def\nF{{n_F^{}}}
\def\nB{{n_B^{}}}
\def\nFp{{n_F^{+}}}
\def\nFm{{n_F^{-}}}
\def\nBp{{n_B^{+}}}
\def\nBm{{n_B^{-}}}

\def\etac{\eta^{}_{corr}}

\def\Deta{\Delta\eta}
\def\Dphi{\Delta\phi}
\def\deta{\delta\eta}
\def\dphi{\delta\phi}
\def\detaF{\delta\eta_{\!F}}
\def\detaB{\delta\eta_{\!B}}
\def\etas{\eta_{sep}}
\def\phis{\phi_{sep}}
\def\etag{\eta_{gap}}

\def\omN{{\omega_N}}

\def\fv{\varphi}

\def\Lam#1{\Lambda(#1)}

\def\nF{{n_F^{}}}
\def\nB{{n_B^{}}}

\def\muF{{\mu_F}}
\def\muB{{\mu_B}}

\def\nF{{n_F^{}}}

\def\nB{{n_B^{}}}

\def\cov{{\textrm{cov}}}

\def\SigFB{{\Sigma(\nF,\nB)}}

\def\Sigmu{{\Sigma (\muF, \muB)}}

\def\Sigk{{\Sigma_{k} (\muF, \muB)}}
\def\Sigmuk{{\Sigma_{k} (\muF, \muB)}}
\def\Lamk{{\Lambda_{k}}}
\def\Lamkk{{\Lambda_{0}^{(k)}}}
\def\etakc{{\eta^{(k)}_{corr}}}
\def\muk{{\mu_{0}^{(k)}}}

\begin{document}
%
\title{Strongly intensive observable between multiplicities\\ in two acceptance windows in a string model}
\author{Evgeny Andronov\thanks{{e.v.andronov@spbu.ru}}
\ and
Vladimir Vechernin
\thanks{{v.vechernin@spbu.ru}}%
}                     
%
%
%
\date{Saint Petersburg State University, 7/9 Universitetskaya nab.,\\ St.Petersburg, 199034 Russia}
%
\maketitle

\abstract{
The strongly intensive observable between multiplicities
in two acceptance windows separated in rapidity and azimuth is calculated
in the model with quark-gluon strings acting as sources.
The dependence of this observable
on the two-particle correlation function of a string, the width of observation windows and
the rapidity gap between them is analyzed.\\
In the case with independent identical strings the model calculation confirms the strongly
intensive character of this observable: it is independent of both the mean number of string
and its fluctuation. For this case the peculiarities of its behaviour for particles with
different electric charges are also analyzed.\\
In the case when the string fusion processes are taken into account and a formation of strings
of a few different types takes place in a collision,
this observable is proved to be equal to a weighted average of its values for different string types.
Unfortunately, in this case through the weight factors this observable becomes
dependent on collision conditions and, strictly speaking, can not be considered
any more as strongly intensive variable.\\
For a comparison the results of the calculation of considered observable with
the PYTHIA event generator are also presented. \\  \ \\
%
      {25.75.Gz} {Particle correlations and fluctuations} \\
      {13.85.Hd} {Inelastic scattering: many-particle final states}  \\
} 
\section{Introduction}
\label{sec-intro}
It is known that the investigations of long range rapidity correlations give the information
about the initial stage of high energy hadronic interactions \cite{Dumitru08}.
So, to find a signature of the string fusion and percolation phenomenon \cite{Biro84,Bialas86,BP92,BP93}
in ultrarelativistic heavy ion collisions
the study of the correlations between multiplicities
in two separated rapidity intervals
(known as the forward-backward multiplicity correlations) was proposed \cite{PRL94}.

Later it was realized \cite{BP00,EPJC04,PPR,YF07-1,YF07-2,TMF15,TMF17}
that the investigations of the forward-backward correlations involving intensive observables
in forward and backward observation windows as, {\it e.g.}, the event-mean transverse momentum,
enable to suppress the contribution of trivial, so-called, "volume" fluctuations
originating from fluctuations in the number of initial sources (strings)
and to obtain more clear signal on the process of string fusion,
compared to usual forward-backward multiplicity correlations.

In the present work, we explore another way to suppress the contribution of "volume" fluctuations,
turning
to the more sophisticated correlation observable.
Basing on the multiplicities $\nF$ and $\nB$ in two separated rapidity windows
we study the properties of the so-called strongly intensive observable
\beq
\label{SigmaFB}
\Sigma(\nF,\nB)\equiv\frac{\av\nF\,\omega_\nB+\av\nB\,\omega_\nF-2\, \cov(\nF\,\nB)}{\av\nF+\av\nB}  \ ,
\eeq
introduced in \cite{GorGaz11}, where
\beq
\label{cov}
\cov(\nF,\nB)\equiv\av{\nF\nB}-\av\nF \av\nB  \ ,
\eeq
and $\omega_\nF$ and $\omega_\nB$ are the corresponding scaled variances of the multiplicities:
\beq
\label{om}
\omega_n\equiv \frac{D_n}{\av n} = \frac{\av {n^2}-\av n^2}{\av n}  \ .
\eeq

In the framework of the model with
color strings as particle emitting sources
we calculate the dependence of the observable (\ref{SigmaFB})
on the string two-particle correlation function,
the width of observation windows and
the rapidity gap between them.
We show that in the case with independent identical strings
the strongly intensive character of this observable is being confirmed:
it depends only on the individual characteristics of a string and
is independent of both
the mean number of strings and its fluctuation.
We also analyze the peculiarities of the behaviour of
the strongly intensive observables between multiplicities
of particles with different electric charges and
as well between multiplicities in two windows separated in rapidity and azimuth.

In the case when the string fusion processes
are taken into account and a formation
of strings of a few different types takes place
we found that,
this observable is equal to a weighted
average of its values for different string types.
Unfortunately, in this case
through the weight factors
the observable becomes dependent on collision conditions
and, strictly speaking, can not be considered
any more as strongly intensive variable.
We also present for a comparison the results of the calculation
of considered observable with PYTHIA event generator.

The paper is organized as follows.
In Section 2 we consider the most simple case
with independent identical strings and symmetric
$2\pi$-azimuth observation windows separated by
a rapidity gap.
In Section 3 we generalize the obtained results for the case
of two acceptance windows separated in rapidity and azimuth.
Section 4 is devoted to the calculations of
the strongly intensive observables between multiplicities
of particles with different electric charges.
In Section 4 the influence of the string fusion processes
is analysed.
Section 5 is devoted to the results obtained
with the PYTHIA event generator.

\section{$\Sigma$ in the model with independent identical strings}
\label{sec-model}
We start our consideration from the simple case
with  symmetric $2\pi$-azimuth observation windows $\detaF=\detaB\equiv\deta$ separated by
a rapidity gap $\etag$, which corresponds to the distance $\Deta=\etag+\deta$ between their centers.
Clear that for symmetric reaction we have
\beq
\label{n}
\av\nF=\av\nB\equiv \av n  \ ,  \hs1
\omega_\nF=\omega_\nB\equiv \omega_n \eeq
and the expression (\ref{SigmaFB}) can be simplified to
\begin{eqnarray}
&&\SigFB=\omn - \cov(\nF,\nB)/\av n=\nonumber\\
&&=\frac{D_n-\cov(\nF,\nB)}{\av n}=  \frac{\av  {n^2}-\av{\nF\nB}}{\av n} \label{Sigma-s}\ .
\end{eqnarray}

In the framework of the model with independent identical strings \cite{PLB00}
we suppose that the number of strings, $N$, fluctuates event by event
around some mean value,  $\av N$, with some scaled variance, $\omN={D_N}/{\av{N}}$.
We expect that the intensive observables should not depend on $\av N$ and
the strongly intensive observables should not depend on both $\av N$  and $\omN$.

The fragmentation of each string contributes to the forward and backward observation rapidity windows,
$\deta_F$, and $\deta_B$,
the $\mu_F$ and $\mu_B$ charged particles correspondingly,
which fluctuate around some mean values, $\av {\mu_F}$ and $\av {\mu_B}$,\
with some scaled variances, $\omega_{\mu_F}={D_{\mu_F}}/{\av{\mu_F}}$
and $\omega_{\mu_B}={D_{\mu_B}}/{\av{\mu_B}}$.
Similarly to (\ref{SigmaFB}) we can formally introduce also
the $\Sigmu$ - the strongly intensive observable between multiplicities,
produced from decay of a single string:
\beq
\label{SigmuFB}
\Sigma(\muF,\muB)\equiv\frac{\av\muF\,\omega_\muB
+\av\muB\,\omega_\muF-2\, \cov(\muF\,\muB)}{\av\muF+\av\muB} \ .
\eeq
For symmetric reaction and symmetric observation windows,
\beq
\label{avmu}
\av {\mu_F}=\av {\mu_B}\equiv\av {\mu}, \hs1 \omega_{\mu_F}=\omega_{\mu_B}\equiv\omega_\mu \ , \eeq
it also can be simplified to
\begin{eqnarray}
&&\Sigmu=\ommu - \cov(\muF,\muB)/\av \mu= \nonumber\\
&&=\frac{D_\mu-\cov(\muF,\muB)}{\av \mu}=
\frac{\av  {\mu^2}-\av{\muF\muB}}{\av \mu} \label{Sigmu-s}\ .
\end{eqnarray}

Clear that in this model
\beq
\label{avn}
\av n=\av N \av{\mu}=\av N \mu_0 \, \deta \ , \hs1 \omega_n=\omega_{\mu}+\omN \av{\mu} \ ,
\eeq
so we see that the $\omega_n$ is intensive, but not strongly intensive observable.
We also supposed the translation invariance in rapidity,
where $\mu_0$ is a distribution density for particles produced from a single string.

For us it is important to remember that the scaled variance, $\omega_n$,
of the number of particles produced in the rapidity interval $\deta$
is determined by the two-particle correlation function $C_2(\eta_1\!-\!\eta_2)$
\cite{CapKrz78,NPA15}:
\beq
\label{omn}
\omega_n=1+ \av n \, I_{FF}  \ ,
\eeq
where
\beq
\label{IFF}
I_{FF}\equiv\frac{1}{\deta^2}  \int_{\deta_F}\!\!\!d\eta_1 \int_{\deta_F} \!\!\!d\eta_2\
C_2(\eta_1\!-\!\eta_2)   \ .
\eeq
Here the two-particle correlation function is defined by the standard way (see, {\it e.g.}, \cite{Voloshin02}):
\begin{equation}
\label{C2def}
C_2(\eta_1,\eta_2)\equiv\frac{\rho_2 (\eta_1,\eta_2)}{\rho(\eta_1) \rho(\eta_2)}-1 \ ,
\end{equation}
and
\begin{equation}
\label{rho12}
\rho (\eta)\equiv\frac{dN_{ch}}{d\eta}
\ , \hs1
\rho_2 (\eta_1, \eta_2)\equiv\frac{d^2N_{ch}}{d\eta_1\,d\eta_2}   \ .
\end{equation}
In the case with a translation invariance in rapidity,
we have the uniform distribution
\begin{equation}
\label{rho1}
\rho (\eta)=\rho_0 \ , \hs1 \rho_2 (\eta_1, \eta_2)=\rho_2 (\eta_1\!-\! \eta_2)
\end{equation}
and the correlation function $C_2(\eta_1\!-\! \eta_2)$ depending only on a difference of rapidities.

Similarly we can also introduce
the two-particle correlation function of a single string
\begin{equation}
\label{Lambda}
\Lambda(\eta_1,\eta_2)\equiv\frac{\lambda_2 (\eta_1,\eta_2)}{\lambda(\eta_1) \lambda(\eta_2)}-1
\end{equation}
for description of the correlation between particles produced from a same string,
where $\lambda(\eta)$ and $\lambda_2 (\eta_1,\eta_2)$ are the corresponding single and
double distributions.
For the translation invariant case
\begin{equation}
\label{rho3}
\lambda (\eta)=\mu_0 \ , \hs1 \lambda_2 (\eta_1, \eta_2)=\lambda_2 (\eta_1\!-\! \eta_2)
\end{equation}
and similarly we have
\beq
\label{omn1}
\omega_\mu=1+ \av\mu\, J_{FF}  \ ,
\eeq
where
\beq
\label{JFF}
J_{FF}\equiv\frac{1}{\deta^2}  \int_{\deta_F}\!\!\!d\eta_1 \int_{\deta_F} \!\!\!d\eta_2\
\Lambda(\eta_1\!-\!\eta_2)   \ ,
\eeq
and the string two-particle correlation function $\Lambda (\eta_1\!-\! \eta_2)$ depends
only on a difference of rapidities.

Note that by formula (\ref{omn1}) we see that the so-called robust variance \cite{Voloshin02}:
\beq
\label{Rob}
R_\mu\equiv\frac{\omega_\mu-1}{\av\mu}= J_{FF}
\eeq
and depends only on a string correlation function $\Lambda(\eta_1\!-\!\eta_2)$.

The similar formulae are valid for the corresponding covariances \cite{NPA15}:
\beq
\label{nFnB}
\frac{\cov(\nF,\nB)}{\av\nF\av\nB} = I_{FB} \ ,
\eeq
where
\beq
\label{IFB}
I_{FB}\equiv\frac{1}{\deta^2}  \int_{\deta_F}\!\!\!d\eta_1 \int_{\deta_B} \!\!\!d\eta_2\
C_2(\eta_1\!-\!\eta_2)   \ ,
\eeq
where the integration over $\eta_1$ and $\eta_2$ are fulfilled now on different
rapidity intervals $\deta_F$ and $\deta_B$ correspondingly. For one string we also have
\beq
\label{muFmuB}
\frac{\cov(\muF,\muB)}{\av\muF\av\muB} = J_{FB} \ ,
\eeq
where
\beq
\label{JFB}
J_{FB}\equiv\frac{1}{\deta^2}  \int_{\deta_F}\!\!\!d\eta_1 \int_{\deta_B} \!\!\!d\eta_2\
\Lambda(\eta_1\!-\!\eta_2)   \ ,
\eeq

In the considered model with independent identical strings
we can express the observable correlation function $C_2(\eta_1,\eta_2)$
through the correlation function of a single string $\Lambda(\eta_1,\eta_2)$
\cite{NPA15}:
\beq
\label{C2Lam}
C_2(\eta_1,\eta_2)=\frac{\Lambda(\eta_1,\eta_2)+\omN}{\av N}
\eeq
What leads immediately to
\beq\label{integrals_homo_nf}
I_{FB}=\frac{J_{FB}+\omN}{\av N}   \ ,
\hs 1
I_{FF}=\frac{J_{FF}+\omN}{\av N}
\eeq
and for symmetric forward-backward windows
\beq\label{omn3}
\omega_n=D_n/\av n= 1+\av\mu\, [J_{FF}+\omN]  \ ,
\eeq
\beq\label{nFnB3}
\cov(\nF,\nB)/\av n= \av\mu\, [J_{FB}+\omN] \ ,
\eeq
We took also into account that $\rho_0=\av N \mu_0$.
Then by (\ref{Sigma-s}), (\ref{omn}) and (\ref{nFnB}) we obtain
\begin{eqnarray}
&&\SigFB=1+\av n\, [I_{FF}-I_{FB}]=\nonumber\\
&&=1+\av\mu\, [J_{FF}-J_{FB}]=\Sigmu  \label{SigmaJnfu} \ ,
\end{eqnarray}
where the $\Sigmu$ is the strongly intensive observable between multiplicities,
produced from decay of a single string, defined by (\ref{SigmuFB}) and (\ref{Sigmu-s}).

By (\ref{SigmaJnfu}) we really see that in the framework of this model
the observable $\SigFB$ is a strongly intensive,
it is independent of both
the mean number of string $\av N$ and its fluctuation $\omN$.
It depends only on the string parameters $\mu_0$, $\Lambda(\eta_1\!-\!\eta_2)$
and the width of observation windows, $\av\mu=\mu_0\deta$.
Whereas the scaled variance $\omn$ is an intensive, but not a strongly intensive observable,
because it is independent on
the mean number of string $\av N$, but through $\omN$ depends on their fluctuation.
We should note that other quantities that characterize correlations between multiplicities
in two windows such as a correlation coefficient \cite{Uhlig78,Derrick86}
or a variance of asymmetry \cite{Back06} are also not strongly intensive and,
therefore, are more sensitive to experimental event selection procedures.

For small observation windows, of a width $\deta\ll\etac$,
where the $\etac$ is the characteristic correlation length for particles produced from the same string,
the formulae (\ref{omn3}-\ref{SigmaJnfu}) takes especially
simple form:
\beq\label{omn_small_nfu}
\omega_n=D_n/\av n= 1+\mu_0\deta\, [\Lambda(0)+\omN] \ , \eeq
\beq\label{nFnB_small_nfu}
\cov(\nF,\nB)/\av n= \mu_0\deta\, [\Lambda(\Deta)+\omN] \ ,
\eeq
\beq\label{Sigma_small_nfu}
\SigFB= 1+\mu_0\deta\, [\Lambda(0)-\Lambda(\Deta)]=\Sigmu   \ ,
\eeq
where $\Deta=\eta_F\!-\!\eta_B$ is a distance between the centers of the forward
an backward observation windows.
From the last formula we see the main properties of the $\SigFB$, which we expect
in this model. Starting from the value 1
it increases with a distance $\Deta$ between the centers of the observation windows,
since the two-particle correlation function of a string $\Lambda(\Deta)$
decrease with $\Deta$.  The extent of the $\Sigma(\Deta)$ increase
with $\Deta$ is proportional to the width
of the observation windows $\deta$.

More detailed description of the $\SigFB$ needs the knowledge of
the two-particle correlation function of a string $\Lambda(\Deta)$.
In paper \cite{NPA15} in the framework of the model with independent identical strings
this function was fitted using the experimental pp ALICE data
on forward-backward correlations between multiplicities
in windows separated in rapidity and azimuth
at three initial energies together with the value of scaled variance of the number of strings $\omN$
(see table \ref{param}):
\begin{eqnarray}
&&\Lam{\Deta,\Dphi}=\Lambda_1 e^{-\frac{|\Deta|}{\eta_1}} e^{-\frac{\Dphi^2}{\fv^2_1}}  +\nonumber\\
&&+\Lambda_2 \left(e^{-\frac{|\Deta-\eta_0|}{\eta_2}}
+ e^{-\frac{|\Deta+\eta_0|}{\eta_2}}\right)  e^{-\frac{(|\Dphi|-\pi)^2}{\fv^2_2}}  \label{Lam_fit}\ ,
\end{eqnarray}
where it was implied that
\beq \label{f_obl}
|\Dphi|\leq\pi  \ .
\eeq
For $|\Dphi| > \pi$ one must periodically extend $\Lam{\Deta,\Dphi}$ to $\Dphi\to\Dphi+2\pi k$.
With such completion the $\Lam{\Deta,\Dphi}$ meets the following requirements:
\begin{eqnarray}
&&\Lam{-\Deta,\Dphi}=\Lam{\Deta ,\Dphi} \ ,\label{Lam_sym1}\\
&&\Lam{\Deta ,-\Dphi}=\Lam{\Deta ,\Dphi} \ , \label{Lam_sym2}\\
&&\Lam{\Deta ,\Dphi+2\pi k}=\Lam{\Deta ,\Dphi} \ . \label{Lam_sym3}
\end{eqnarray}

\begin{table}[!tb]
 \caption[dummy]{\label{param}
The value of the parameters in formula (\ref{Lam_fit}) \cite{NPA15}
for the two-particle correlation function of a string $\Lambda(\Deta, \Dphi)$
fitted by the experimental pp ALICE data
on forward-backward correlations between multiplicities
in windows separated in rapidity and azimuth
at three initial energies \cite{ALICE15}
together with the value of scaled variance of the number of strings $\omN$.
}
  \centering
\begin{tabular}{|c|c|c|c|c|}
 \hline
 \multicolumn{2}{|c|}{$\sqrt{s}$,\  TeV}&0.9&2.76&7.0 \\
 \hline  \hline
 LRC &$\mu_0\omega_N$&0.7 & 1.4 &2.1\\
 \hline \hline
 &$\mu_0\Lambda_1$&1.5 & 1.9 & 2.3 \\
 &$\eta_1$&0.75 &0.75&0.75 \\
 &$\phi_1$&1.2 &1.15&1.1 \\
 \cline{2-5}
 SRC&$\mu_0\Lambda_2$&0.4 &0.4&0.4 \\
 &$\eta_2$&2.0 &2.0&2.0 \\
 &$\phi_2$&1.7 &1.7&1.7 \\
 \cline{2-5}
 &$\eta_0$&0.9 &0.9&0.9 \\
 \hline
\end{tabular}
\end{table}

Recall that the comparison of the model with experimental data in \cite{NPA15} enables
to fix only the product of the parameters $\mu_0\Lambda_1$, $\mu_0\Lambda_2$ and $\mu_0\omN$.

\begin{figure}[!tb]
\centering
\includegraphics[width=0.5\textwidth,angle=-90]{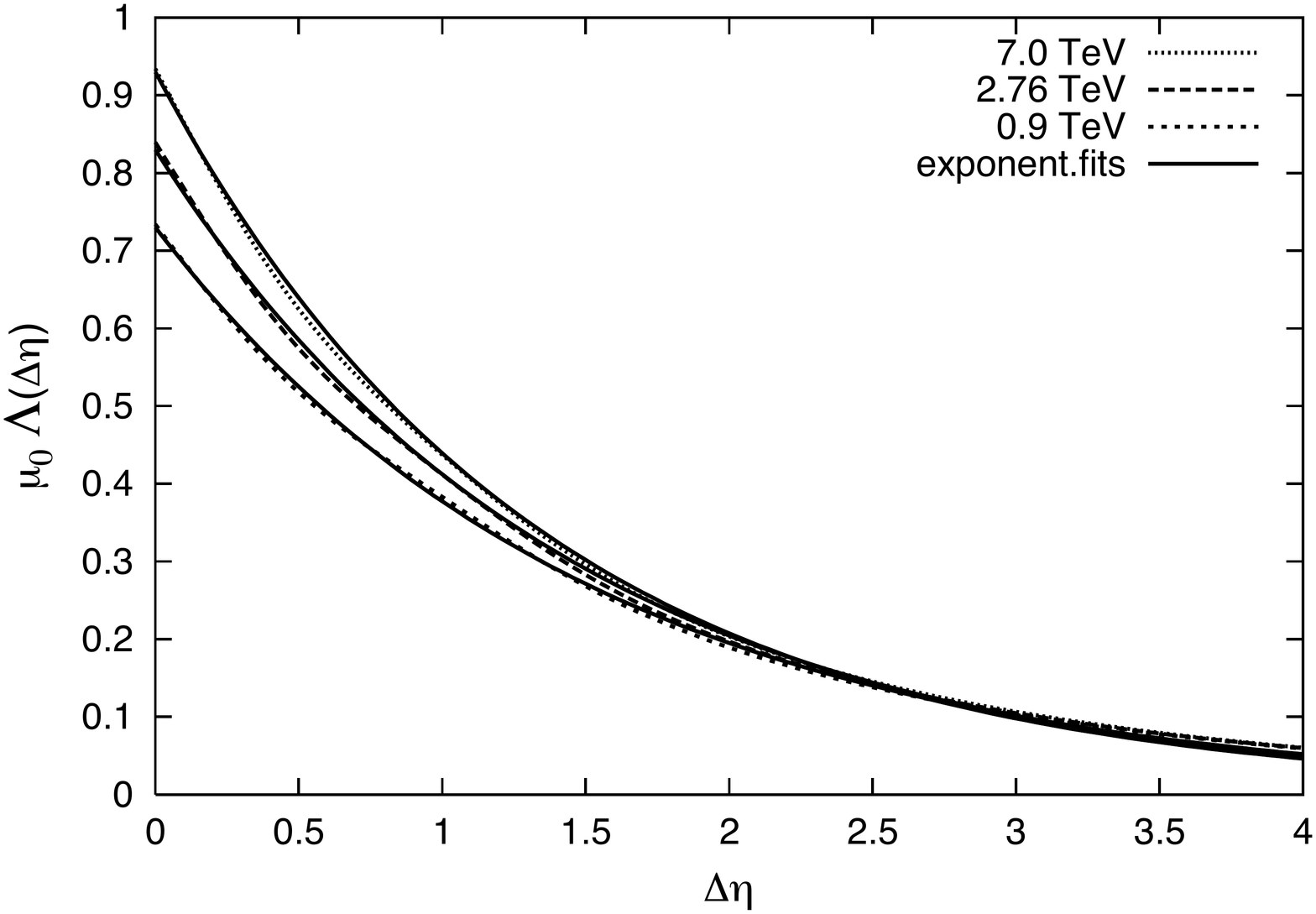}
\caption{\label{Lam-exp}
The two-particle correlation function of a string $\Lambda(\Deta)$ (integrated over azimuth)
obtained by a fitting \cite{NPA15} of the experimental pp ALICE data \cite{ALICE15}
on forward-backward correlations between multiplicities at three initial energies: 0.9, 2.76 and 7 TeV
(the dashed lines) and the corresponding exponential fits (\ref{Lam_exp})  (solid lines) with the parameters
presented in table \ref{exp-par}.
}
\end{figure}

Our two-particle correlation functions (\ref{C2def}) and (\ref{Lambda})
defined for $2\pi$-azimuth observation windows
can be obtained by simple integration over azimuth:
\begin{eqnarray}
&&C_2(\Deta) =
\frac{1}{\pi}\int_{0}^{\pi} \!\!   C_2(\Deta,\Delta\phi) \,  d\Delta\phi \ , \label{int_azim1}\\
&&\Lambda(\Deta) =
\frac{1}{\pi}\int_{0}^{\pi} \!\!   \Lambda(\Deta,\Delta\fv) \, d\Delta\fv    \ .\label{int_azim2}
\end{eqnarray}
So by integration of the fit (\ref{Lam_fit}) we find the $\Lambda(\Deta)$ presented in Fig.~\ref{Lam-exp}.
The obtained dependencies in this figure for three initial energies are well
approximated by the exponent
\beq \label{Lam_exp}
\Lambda(\Deta)=\Lambda_0 \exp(-{|\Deta|}/{\etac})  \ ,
\eeq
with the parameters presented in table~\ref{exp-par}.
We see that the correlation length, $\etac$, decreases with the increase of collision energy.
This can be interpreted as a signal of an increase with energy of the admixture
of strings of a new type - the fused strings in pp collisions (see below).

\begin{table}[!tb]
 \caption[dummy]{\label{exp-par}
The value of the parameters in formula (\ref{Lam_exp})
for the two-particle correlation function of a string $\Lambda(\Deta)$
obtained by a fitting \cite{NPA15} of the experimental pp ALICE data \cite{ALICE15}
on forward-backward correlations between multiplicities at three initial energies
(see Fig.~\ref{Lam-exp}).
}
  \centering
\begin{tabular}{|c|c|c|c|}
 \hline
$\sqrt{s}$,\  TeV&0.9&2.76&7.0 \\
 \hline  \hline
 $\mu_0\Lambda_0$&0.73 & 0.83 &0.93\\
 $\etac$                    &1.52 & 1.43 &1.33\\
 \hline
\end{tabular}
\end{table}


\begin{figure}[!tb]
\centering{
\includegraphics[width=0.5\textwidth,angle=0]{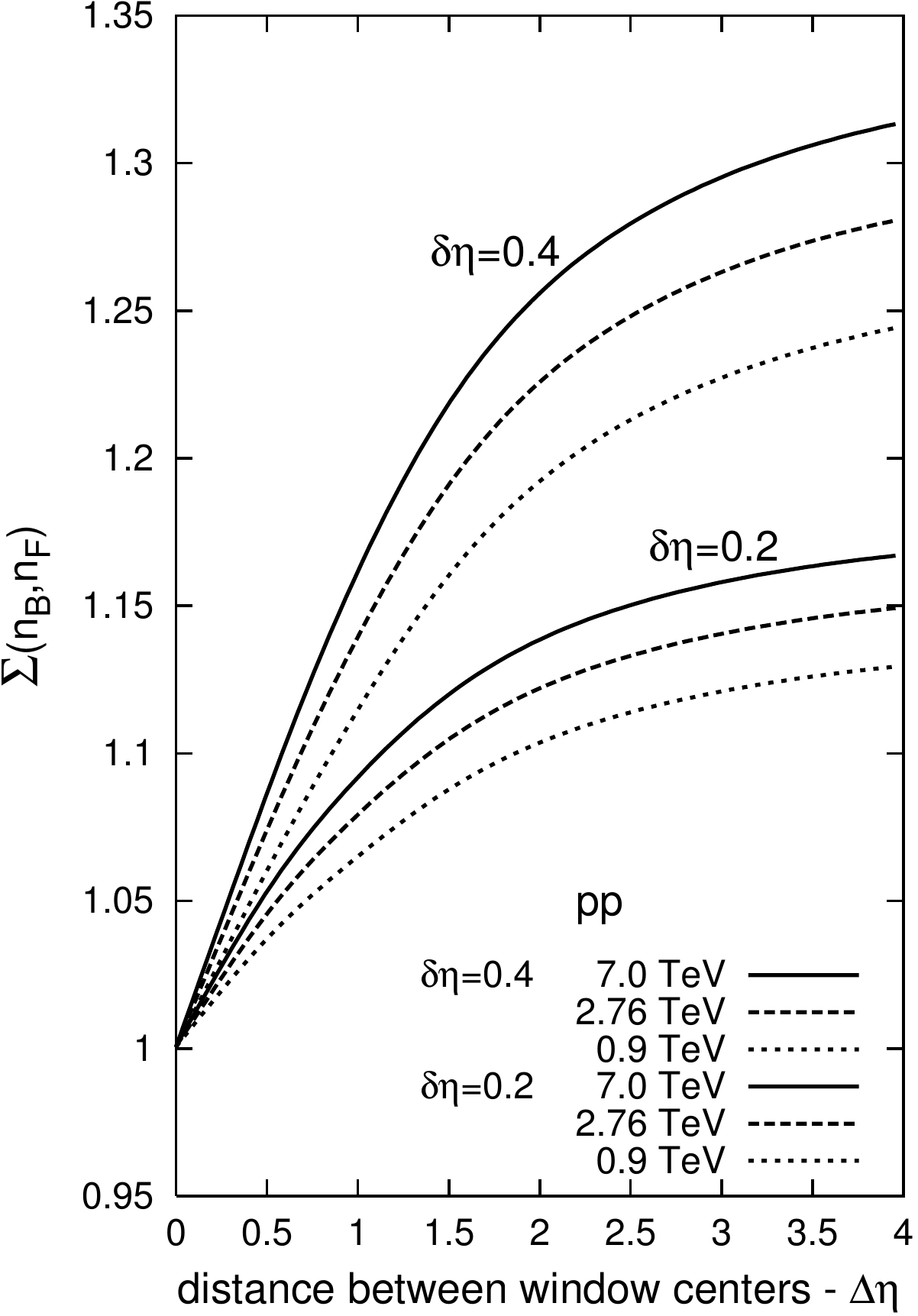}
}
\caption{\label{Sigma-pp}
The strongly intensive observable, $\SigFB$, between multiplicities
in two small pseudorapidity windows (of the width $\deta=$ 0.2 and 0.4)
as a function of the distance between window centers, $\Deta$, calculated
in the model with independent identical strings using
the two-particle correlation function of a string $\Lambda(\Deta)$ (see Fig.\ref{Lam-exp})
obtained by a fitting \cite{NPA15} of the experimental pp ALICE data \cite{ALICE15}
on forward-backward correlations between multiplicities at three initial energies: 0.9, 2.76 and 7 TeV.
}
\end{figure}

The results of the calculation of the strongly intensive observable $\SigFB$
by formulae (\ref{SigmaJnfu}) with this two-particle correlation function
for two width of the observation windows $\deta=0.2$ and $0.4$
are presented in Fig.~\ref{Sigma-pp} for three initial energies: 0.9, 2.76 and 7 TeV.

By formulae (\ref{Sigma-s}) and (\ref{omn_small_nfu}-\ref{Sigma_small_nfu})
we can understand the behaviour of $\SigFB$ in this figure as follows.
The formula (\ref{Sigma-s}) shows that
for symmetric reaction and symmetric observation windows
the $\SigFB$ is proportional to the difference
between the variance $D_{n_F}$ and the covariance $\cov(\nF,\nB)$.
It is important to remember that the value both of them
are determined by the string two-particle correlation function $\Lambda$
and the scaled variance in the number of strings $\omN$
(see formulae (\ref{omn_small_nfu}) and (\ref{nFnB_small_nfu})).
In particular in the absence of correlations between particles
produced from a given source
the multiplicity distribution
from such source will be poissonian ($\ommu=1$, see formula (\ref{omn1})).

First of all we see that in $\SigFB$,
which by (\ref{Sigma-s}) is a difference between $D_n/\av n$ and $\cov(\nF,\nB)/\av n$,
the contributions from the variance in the number of strings, $\omN$,
are being mutually canceled (see formulae (\ref{omn3}), (\ref{nFnB3})
or (\ref{omn_small_nfu}), (\ref{nFnB_small_nfu})),
what reflects the strongly intensive character of the quantity.
Moreover by (\ref{omn_small_nfu}) we see that
at small values of the distance between observation windows $\Deta\ll\etac$
the contribution, $\mu_0\deta\Lambda(0)$, of the two-particle correlations to $\omn$
is being compensated by their contribution, $\mu_0\deta\Lambda(\Deta)$, to $\cov(\nF,\nB)/\av n$
and the $\SigFB$ is equal to 1.

At large distances between observation windows $\Deta\gg\etac$,
by formula (\ref{Lam_exp}), the two-particle correlation function of a  string, $\Lambda(\Deta)$,
goes to zero and
the $\SigFB$ saturates to $\ommu=1+\mu_0\deta\Lambda(0)$. So we have
\begin{eqnarray}
&&\SigFB\to1 \hs{0.5} \textrm{at}  \hs{0.5} \Deta\ll\etac   \ , \label{limit1-1}\\
&&\SigFB\to\ommu \hs{0.5} \textrm{at}  \hs{0.5} \Deta\gg\etac   \label{limit1-2}\ .
\end{eqnarray}
Note that $\ommu$ increases with the width, $\deta$, of the observation windows,  (\ref{omn_small_nfu}).

\begin{figure}[!tb]
\centering{
\includegraphics[width=70mm,angle=0]{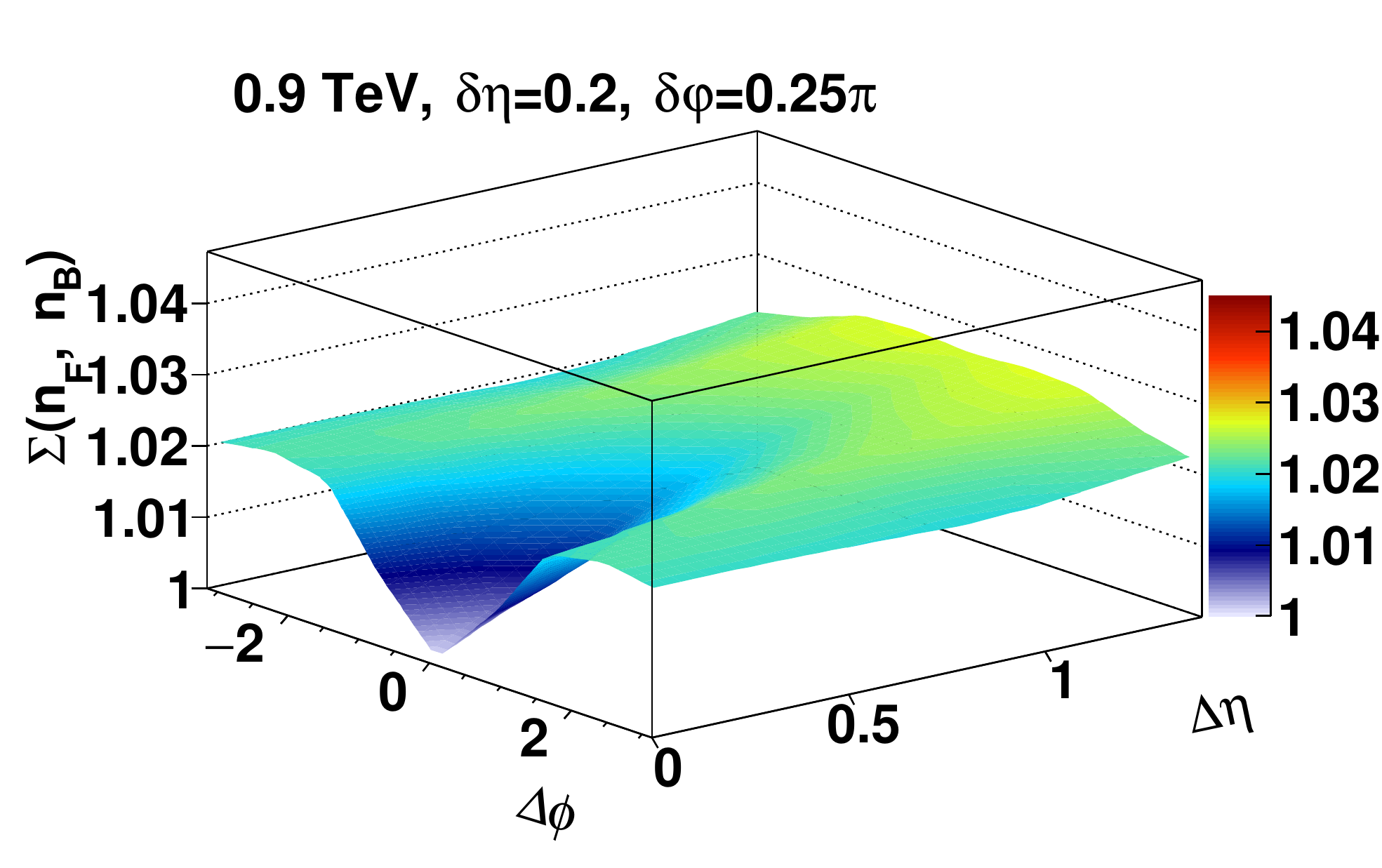}
}
\centering{
\includegraphics[width=70mm,angle=0]{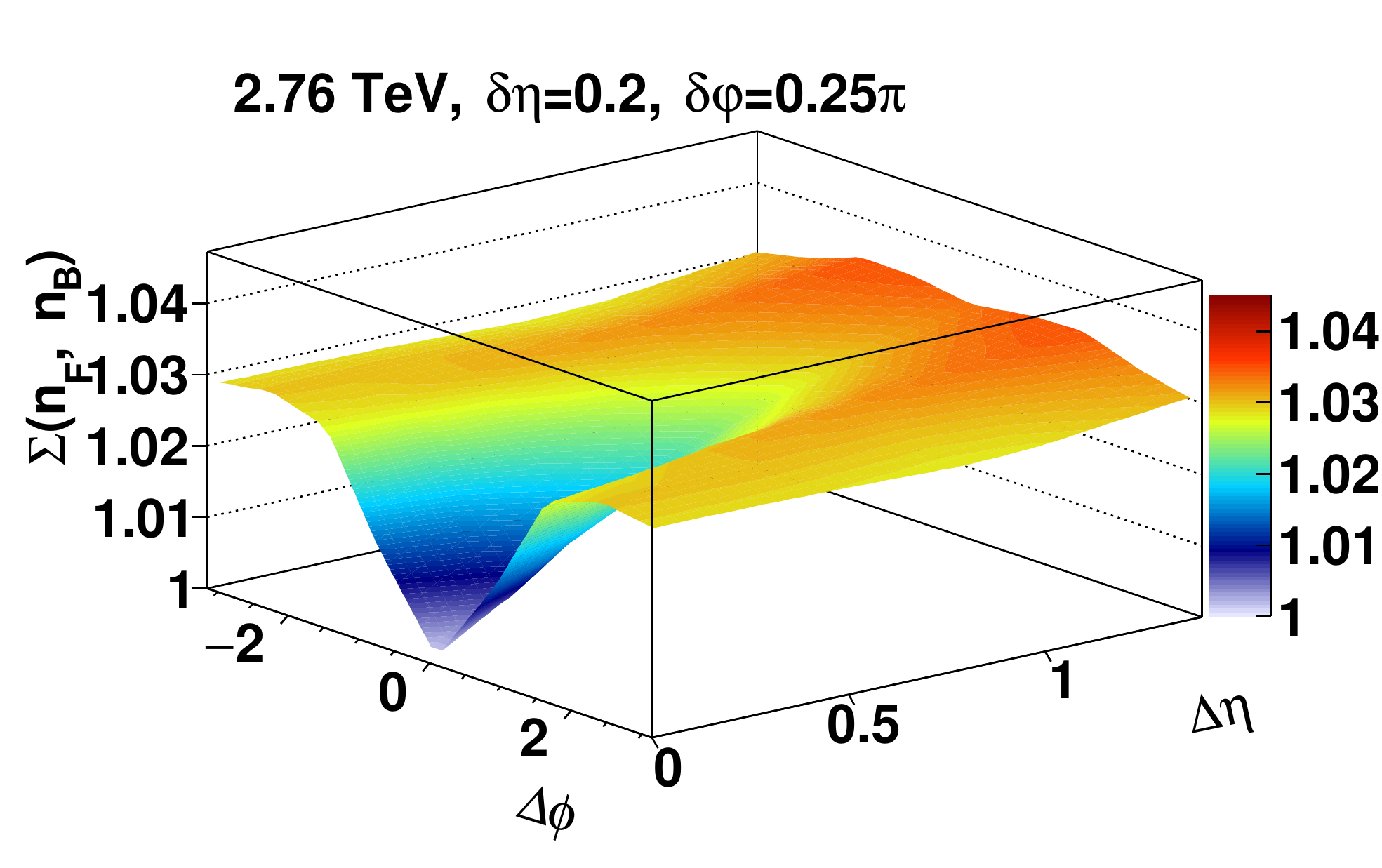}
}
\centering{
\includegraphics[width=70mm,angle=0]{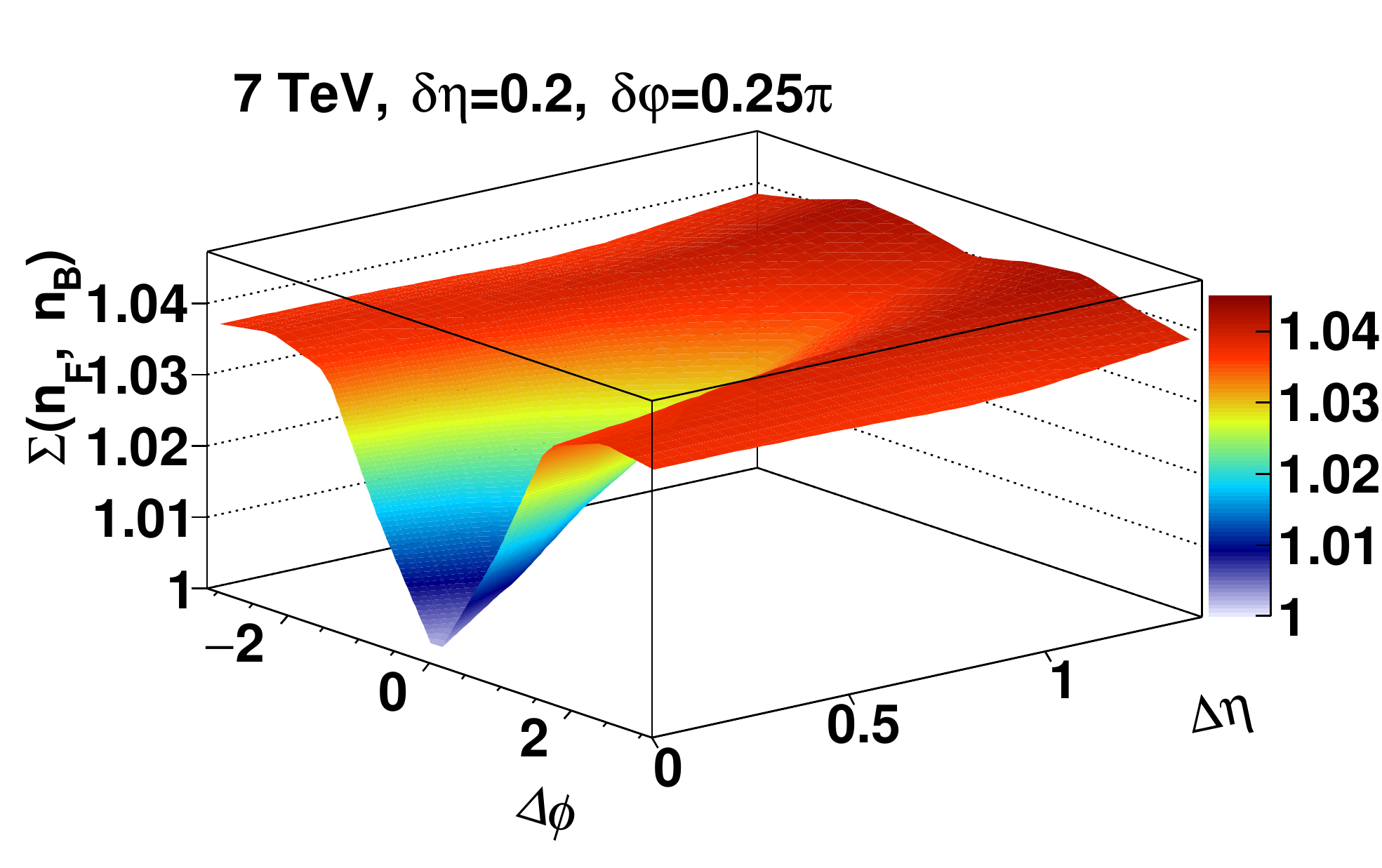}
}
\caption{\label{Sigma-dd1}
The strongly intensive observable, $\SigFB$, between multiplicities
in two windows of the width $\deta=$ 0.2 and $\dphi=\pi/4$
as a function of the distance between window centers $\Deta$ in rapidity and $\Dphi$ in azimuth,
calculated
in the model with independent identical strings using
the two-particle correlation function of a string $\Lambda(\Deta,\Dphi)$ (\ref{Lam_fit})
obtained by a fitting \cite{NPA15} of the experimental pp ALICE data \cite{ALICE15}
on forward-backward correlations between multiplicities at three initial energies: 0.9, 2.76 and 7 TeV
in ALICE TPC pseudorapidity acceptance.
}
\end{figure}

In Fig.~\ref{Sigma-pp} we see also
some general increase of the $\SigFB$ with initial energy,
below in Section \ref{sec-fusion} we will show that
in the framework of the string model it can be interpreted
as a signal of an increase with energy of the admixture
of strings of a new type - the fused strings in pp collisions.


\section{$\Sigma$ for windows separated in rapidity and azimuth}
\label{sec-azimuth}

\begin{figure}[!tb]
\centering{
\includegraphics[width=70mm,angle=0]{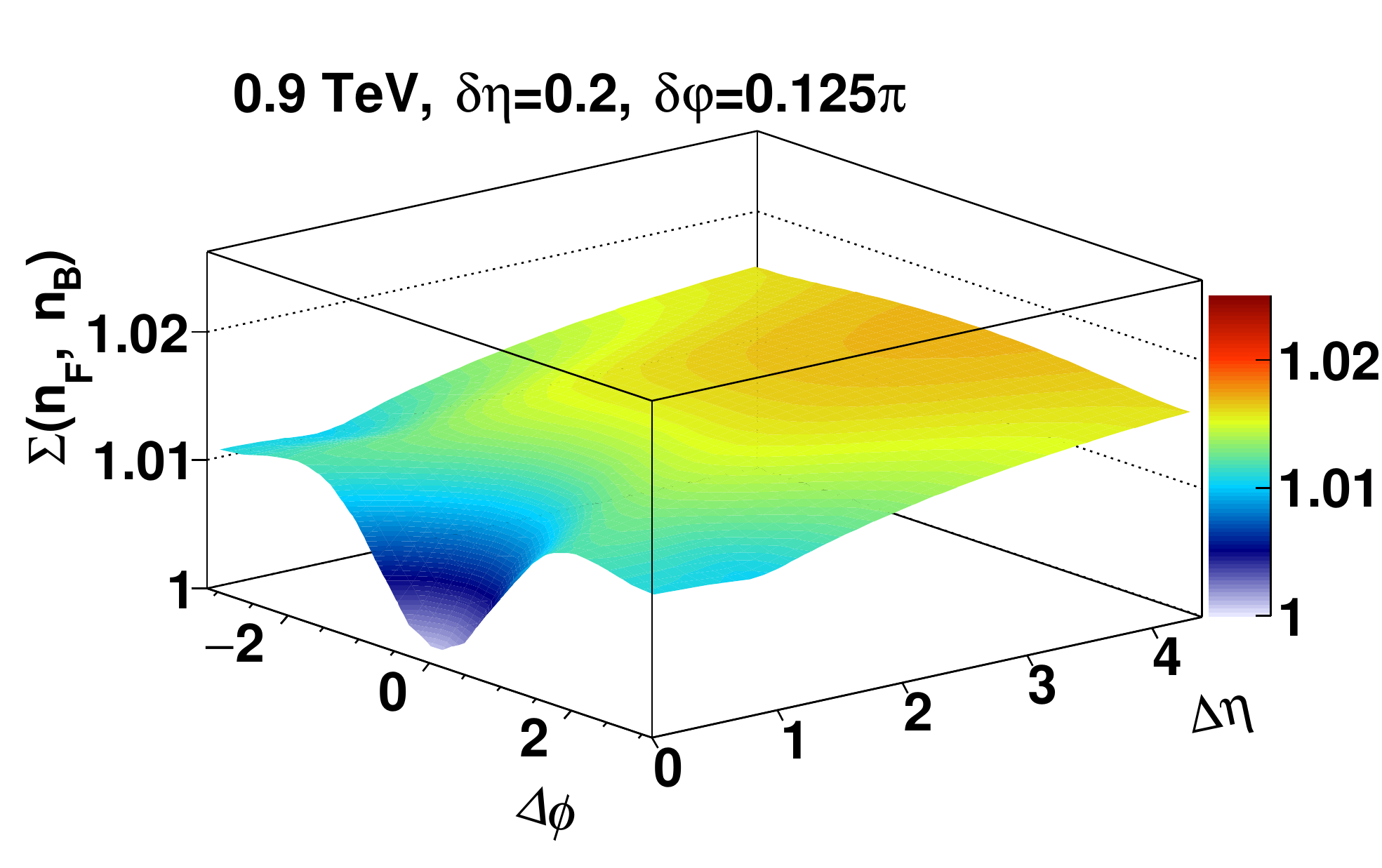}
}
\centering{
\includegraphics[width=70mm,angle=0]{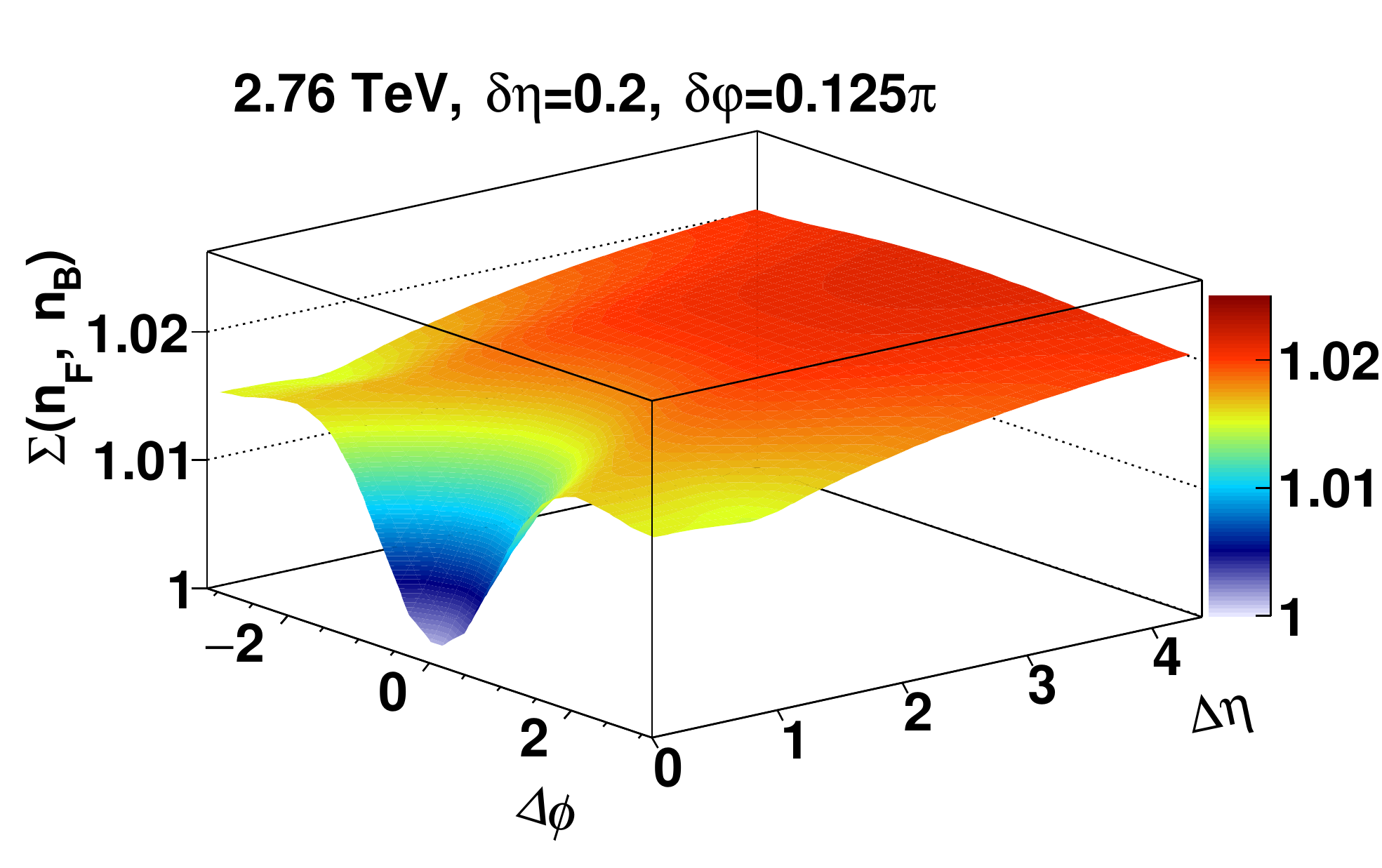}
}
\centering{
\includegraphics[width=70mm,angle=0]{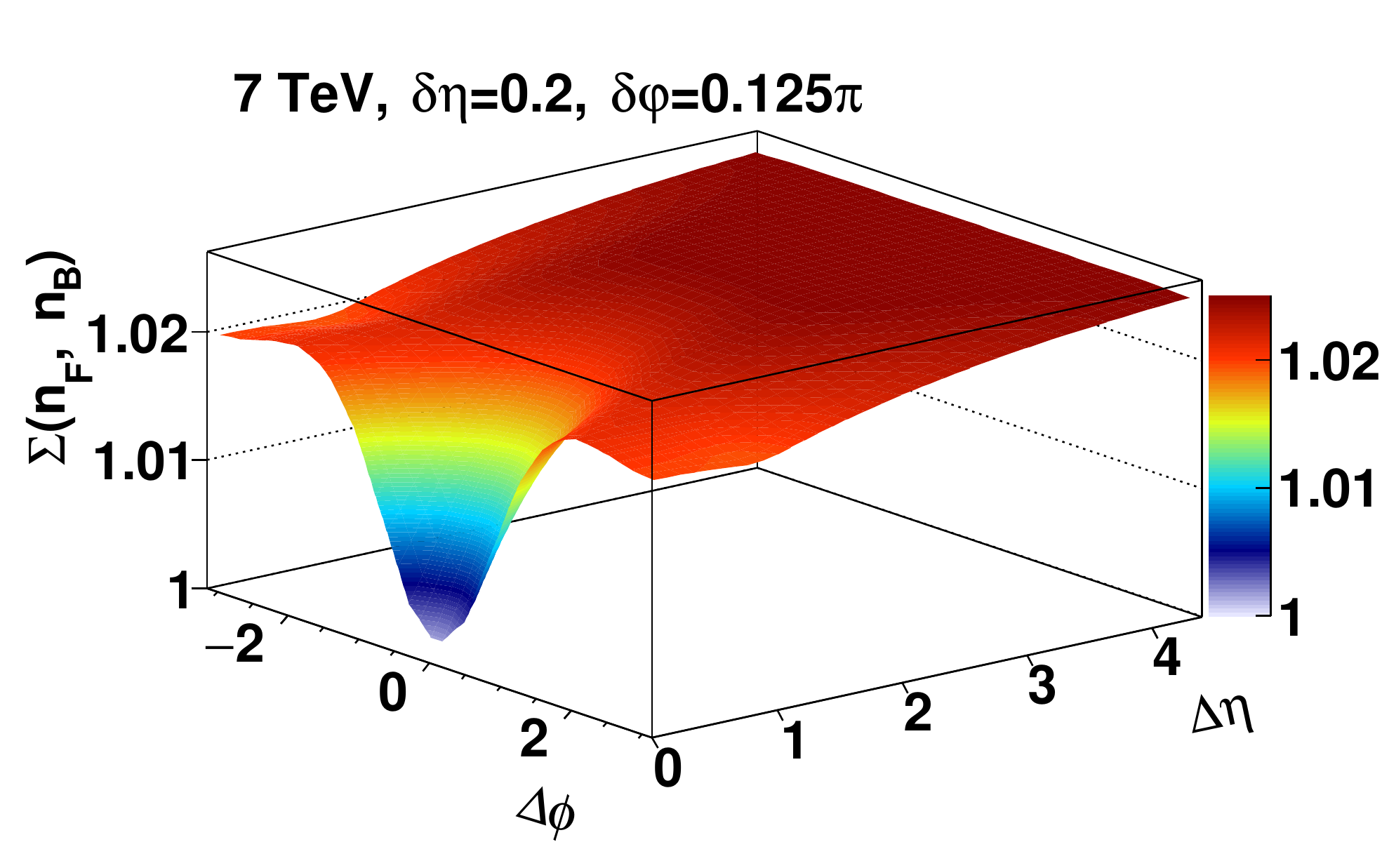}
}
\caption{\label{Sigma-dd2}
The same as in Fig.\ref{Sigma-dd1}, but with $\pi/8$ azimuth window width and
extrapolated to a wider interval of the separation between windows in rapidity:
the $\Deta$ is upto 4 rapidity units.
}
\end{figure}

All results obtained in Section \ref{sec-model}
can be easily extended to the case of
the strongly intensive observable $\SigFB$ between multiplicities
in two acceptance windows separated both in rapidity and azimuth.

In particular for symmetric reaction and symmetric small observation windows
of the width $\deta$ in rapidity and $\dphi$ in azimuth
with the separation between their centers $\etas=\Deta$ and $\phis=\Dphi$
we find in the model with independent identical strings:
\beq\label{Sigmadd}
\SigFB
=1+ \frac{\deta\,\dphi}{2\pi} \mu_0\ [\Lambda(0,0)-\Lambda(\Deta,\Dphi)]      \ ,
\eeq
which is a generalization of the formula (\ref{Sigma_small_nfu}).

If we use now again for
the two-particle correlation function of a string $\Lambda(\Deta,\Dphi)$ the approximation
(\ref{Lam_fit}) suggested in paper \cite{NPA15}
with the parameters (see table \ref{param}) fitted by the experimental pp ALICE data \cite{ALICE15}
on forward-backward correlations between multiplicities
in windows separated in rapidity and azimuth
at three initial energies, then we find the behaviour of $\SigFB$,
presented in Fig.\ref{Sigma-dd1} in ALICE TPC acceptance and
in Fig.\ref{Sigma-dd2} extrapolated to a wider rapidity interval.

The explanation of this behaviour of $\SigFB$
on the base of the formula \ref{Sigmadd} is
absolutely the same as in the end of the Section \ref{sec-model}.

\section{$\Sigma$ with charges}
\label{sec-charge}
In Section \ref{sec-model} we have introduced the strongly intensive observable $\Sigma$ based on
multiplicities of the all charged hadrons measured in two pseudorapidity intervals.
Now we consider different combinations of electric charges in these windows
and similarly to formula (\ref{SigmaFB})
we define $\Sigma(n_{F}^{+},n_{B}^{+})$, $\Sigma(n_{F}^{-},n_{B}^{-})$, $\Sigma(n_{F}^{+},n_{B}^{-})$
and $\Sigma(n_{F}^{-},n_{B}^{+})$.

For a symmetric reaction and symmetric observation windows we have
\beq
\label{sym}
\av\nFp=\av\nBp\equiv\av {n^+}   \ ,  \hs{0.5}
\omega_{\nFp}=\omega_{\nBp}\equiv \omega_{n^+}
\eeq
and the same for $n^-$.  In this case we have also
\begin{eqnarray}
&&\cov(\nFp,\nFm)=\cov(\nBp,\nBm) \ ,  \label{symcov-1}\\
&&\cov(\nFp,\nBm)=\cov(\nFm,\nBp) \ , \label{symcov-2}
\end{eqnarray}
and the definitions can be reduced to
\beq\label{Sigma-plusplus-def}
\Sigma(n_{F}^{+},n_{B}^{+})=\omega_{n^{+}}-\frac{\cov(n_{F}^{+},n_{B}^{+})}{\langle n^{+}\rangle} \ ,
\eeq
\beq\label{Sigma-minusminus-def}
\Sigma(n_{F}^{-},n_{B}^{-})=\omega_{n^{-}}-\frac{\cov(n_{F}^{-},n_{B}^{-})}{\langle n^{-}\rangle} \ ,
\eeq
\begin{eqnarray}
&&\Sigma(n_{F}^{+},n_{B}^{-})=\Sigma(n_{F}^{-},n_{B}^{+})=\nonumber\\
&&=\frac{\av {n^+} \omega_{n^-} + \av {n^-} \omega_{n^+}-2\,\cov(\nFp,\nBm)}
{\langle n \rangle}  \ .\label{charges}
\end{eqnarray}
We can also introduce an additional strongly intensive observable
that measures correlation between multiplicities
of different charges in the same window \cite{Andronov16}:
\begin{eqnarray}
&&\Sigma(n_{F}^{+},n_{F}^{-})=\Sigma(n_{B}^{+},n_{B}^{-})=\nonumber\\
&&=\frac{\av {n^+} \omega_{n^-} + \av {n^-} \omega_{n^+}-2\,\cov(\nFp,\nFm)}
{\langle n \rangle}  \ .\label{charges-same}
\end{eqnarray}
By expanding $n_F=n_{F}^{+}+n_{F}^{-}$ and $n_B=n_{B}^{+}+n_{B}^{-}$ in (\ref{Sigma-s})
and taking into account (\ref{Sigma-plusplus-def})-(\ref{charges-same}) we find the following elegant relation:
\begin{eqnarray}
&&\SigFB =
\frac{\av{n^+}}{\av n} \Sigma(n_{F}^{+},n_{B}^{+})
+\frac{\av{n^-}}{\av n} \Sigma(n_{F}^{-},n_{B}^{-})+\nonumber\\
&& +\Sigma(n_{F}^{+},n_{B}^{-})
- \Sigma(n_{F}^{+},n_{F}^{-})\ .\label{Sigma-relation}
\end{eqnarray}

This relation can be further simplified in case of
charge symmetry, when
\beq
\label{sym-ch}
\av{n^+}=\av{n^-}=\av{n}/2   \ ,  
\eeq
and
$$
\omega_{n^+}=\omega_{n^-}   \ ,   \hs{1}
\cov(\nFp,\nBp)=\cov(\nFm,\nBm)  \ ,
$$
what is a very good approximation for mid-rapidity region at LHC collision energies.
In this case we have
\beq\label{Sigma-final-relation}
\SigFB =\Sigma(n_{F}^{+},n_{B}^{+}) +\Sigma(n_{F}^{+},n_{B}^{-}) - \Sigma(n_{F}^{+},n_{F}^{-})\ .
\eeq

In order to calculate charge-dependent strongly intensive observables in the model
of independent strings we have to define corresponding one- and two-particle distributions
describing decay properties of a source. For the charge symmetry case we have:
\begin{eqnarray}
&&\lambda^{+}(\eta)=\lambda^{-}(\eta)=\frac{1}{2}\lambda(\eta)   \ , \label{l1-charge}\\
&&\Lambda^{++}(\eta_{1},\eta_{2})=\Lambda^{--}(\eta_{1},\eta_{2}) \ , \label{l2-charge-1}\\
&&\Lambda^{+-}(\eta_{1},\eta_{2})=\Lambda^{-+}(\eta_{1},\eta_{2}) \ , \label{l2-charge-2}
\end{eqnarray}
\beq \label{L-charge}
\Lambda(\eta_{1},\eta_{2})=\frac{1}{2}\left(\Lambda^{++}(\eta_{1},\eta_{2})+
\Lambda^{+-}(\eta_{1},\eta_{2})\right) \ .
\eeq
Using the translation invariance in rapidity it is again conveniently to define the following quantities:
\begin{eqnarray}
&&J_{FF}^{++}\equiv\frac{1}{\deta^2}  \int_{\deta_F}\!\!\!d\eta_1 \int_{\deta_F} \!\!\!d\eta_2\
\Lambda^{++}(\eta_1\!-\!\eta_2) \ ,  \label{jffplpl}\\
&&J_{FF}^{+-}\equiv\frac{1}{\deta^2}  \int_{\deta_F}\!\!\!d\eta_1 \int_{\deta_F} \!\!\!d\eta_2\
\Lambda^{+-}(\eta_1\!-\!\eta_2)   \ , \\
&& J_{FB}^{++}\equiv\frac{1}{\deta^2}  \int_{\deta_F}\!\!\!d\eta_1 \int_{\deta_B} \!\!\!d\eta_2\
\Lambda^{++}(\eta_1\!-\!\eta_2) \ , \\
&& J_{FB}^{+-}\equiv\frac{1}{\deta^2}  \int_{\deta_F}\!\!\!d\eta_1 \int_{\deta_B} \!\!\!d\eta_2\
\Lambda^{+-}(\eta_1\!-\!\eta_2)   \ . \label{jfbplpl}
\end{eqnarray}
Then by (\ref{JFF}), (\ref{JFB}), (\ref{l1-charge}), (\ref{L-charge}) and (\ref{jffplpl}-\ref{jfbplpl}) we have
\begin{eqnarray}
&&\langle \mu^{+}\rangle = \langle \mu^{-} \rangle = \frac{1}{2} \langle \mu \rangle
=\frac{1}{2}\mu_{0}\delta\eta \ , \label{mu-new}\\
&&J_{FF}=\frac{1}{2}\left(J_{FF}^{++}+J_{FF}^{+-}\right) \ , \label{jff-new}\\
&& J_{FB}=\frac{1}{2}\left(J_{FB}^{++}+J_{FB}^{+-}\right) \ .
\end{eqnarray}
What leads to the following relations:
\beq \label{sigma-plusplus}
\Sigma(n^{+}_{F},n^{+}_{B})=1+\langle\mu^{+}\rangle(J_{FF}^{++}-J_{FB}^{++}) \ ,
\eeq
\beq \label{sigma-plusminus}
\Sigma(n^{+}_{F},n^{-}_{B})=1+\langle\mu^{+}\rangle(J_{FF}^{++}-J_{FB}^{+-}) \ .
\eeq
\beq \label{sigma-plusminus-null}
\Sigma(n^{+}_{F},n^{-}_{F})=1+\langle\mu^{+}\rangle(J_{FF}^{++}-J_{FF}^{+-}) \ ,
\eeq
or in the case of small windows:
\beq \label{sigma-plusplus-small}
\Sigma(n^{+}_{F},n^{+}_{B})=1+\frac{1}{2}\mu_{0}\delta\eta[\Lambda^{++}(0)-\Lambda^{++}(\Delta\eta)] \ ,
\eeq
\beq \label{sigma-plusminus-small}
\Sigma(n^{+}_{F},n^{-}_{B})=1+\frac{1}{2}\mu_{0}\delta\eta[\Lambda^{++}(0)-\Lambda^{+-}(\Delta\eta)] \ ,
\eeq
\beq \label{sigma-plusminus-null-small}
\Sigma(n^{+}_{F},n^{-}_{F})=1+\frac{1}{2}\mu_{0}\delta\eta[\Lambda^{++}(0)-\Lambda^{+-}(0)] \ .
\eeq
We see that as expected $\Sigma(n^{+}_{F},n^{+}_{B})\to1$ at $\Deta\to0$, however
the $\Sigma(n^{+}_{F},n^{-}_{B})$ tends to be equal to
$\Sigma(n^{+}_{F},n^{-}_{F})$ in this limit, which is not necessarily equal to $1$.
To deduce how $\Sigma(n^{+}_{F},n^{-}_{B})$ behaves at small $\Deta$ one need additional input from experiments.

Looking at relations (\ref{sigma-plusplus-small}-\ref{sigma-plusminus-null-small}) one can immediately
notice certain similarities with charge-dependent correlations measured via
the so-called balance function $B\left(\Delta \eta\right)$~\cite{ALICE16}.
In this paper the balance function is defined to be proportional to the difference between unlike-sign
and like-sign two-particle correlations functions:
\beq \label{balance-def}
B(\Deta,\Dphi)\equiv\frac{1}{2}[C_{+-}+C_{-+}-C_{++}-C_{--}],
\eeq
or simply
\beq \label{balance-def-1}
B(\Deta,\Dphi)=C_{+-}-C_{++} \ .
\eeq
The last equation exploits the charge symmetry in mid-rapidities at LHC energies.
From (\ref{C2Lam}) we expect that the $B(\Delta\eta)$ is proportional to
$\Lambda^{+-}(\Delta\eta)\!-\!\Lambda^{++}(\Delta\eta)$,
i.e. to $\Sigma(n^{+}_{F},n^{-}_{B})\!-\!\Sigma(n^{+}_{F},n^{+}_{B})$.
Really, taking into account the normalization of the two-particle correlations functions,
used in paper~\cite{ALICE16}, we find
\beq \label{balance-proj}
B^{proj}(\Deta)=\frac{1}{4}\mu_0[\Lambda^{+-}(\Deta)-\Lambda^{++}(\Deta)]  \ .
\eeq
Here following \cite{ALICE16} the pseudorapidity dependence of balance function is defined as a projection
of two-dimensional $B\left(\Delta\eta,\Delta\phi\right)$:
\beq \label{balance-proj-def}
B^{proj}\left(\Delta\eta\right) \equiv \int_{-\frac{\pi}{2}}^{\frac{3\pi}{2}} B(\Deta,\Dphi) \, d\Dphi   \ .
\eeq
Note that as it is pointed out before the subsection 6.1.1 in the paper \cite{ALICE16}
the results for $B^{proj}(\Deta)$ will be two times larger if calculating the two-particle correlation functions,
entering the definition of balance function (\ref{balance-def-1}),
one will not impose the requirement that the transverse momentum of the
"trigger" particle must be higher than the "associated" one,
{\it i.e.} we would have coefficient 1/2 in (\ref{balance-proj}) instead of 1/4
for the balance function normalized in such a way.

So, by fitting the experimental pp ALICE data on balance functions~\cite{ALICE16} and keeping in mind
relation (\ref{L-charge}) one can extract parameters of unlike-sign and like-sign correlation functions
$\Lambda^{+-}\left(\Delta\eta\right)$ and $\Lambda^{++}\left(\Delta\eta\right)$, and, in turn,
predict $\Delta\eta$ dependencies of the variables
$\Sigma(n^{+}_{F},n^{+}_{B})$ and $\Sigma(n^{+}_{F},n^{-}_{B})$.

For better data fitting we have to take into account the HBT correlations
for like-sign two-particle correlation function and
complement the form of parametrization (\ref{Lam_exp}):
\begin{eqnarray}
&&\Lambda^{++}(\Deta)=\Lambda_0^{++} \exp\left(-{|\Deta|}/{{\eta}^{++}}\right) +\nonumber\\
&&+ \Lambda_0^{HBT} \exp\left[-{\left({\Deta}/{{\eta}^{HBT}}\right)}^{2}\right]\ . \label{Lam_exp_pp}
\end{eqnarray}
We keep it unmodified for unlike-sign correlations\footnote{In order to take into account decays of neutral resonances one need to add a characteristic contribution to the unlike-sign correlations function. We postpone this modification for future research.}
\beq \label{Lam_exp_pm}
\Lambda^{+-}(\Deta)=\Lambda_0^{+-} \exp\left(-{|\Deta|}/{{\eta}^{+-}}\right)  \ .
\eeq
Here, we assumed also that HBT-correlations appears only for pairs of particles originating from the same string.

We perform simultaneous fitting of the experimental data on pp collisions at
7 TeV for balance functions~\cite{ALICE16}
and of $\mu_{0}\Lambda\left(\Delta\eta\right)$ extracted from forward-backward
correlations~\cite{ALICE15}
(see Fig.~\ref{Lam-exp}). As the forward-backward correlations were measured experimentally
for minimum bias pp events, the results on balance functions for 70-80$\%$ pp centrality class
were selected, assuming that the minimum bias is dominated by the 'peripheral' collisions.

Moreover, in order to compensate the more narrow transverse momentum interval taking into account
in F-B correlations measurements compared to balance function investigations
($(0.3;1.5)$ GeV/c in \cite{ALICE15} and $(0.2;2)$ GeV/c in \cite{ALICE16})
we multiply extracted $\mu_{0}\Lambda\left(\Delta\eta\right)$
by a correction factor $c_{cor}$=1.28 that was estimated in the PYTHIA model
\cite{Sjostrand15,Sjostrand06} as a ratio of mean multiplicities at mid-rapidity
for corresponding $p_{T}$ intervals.

\begin{figure}[!tb]
\centering{
\includegraphics[width=\textwidth,angle=0, trim={0 0.4cm 0 0.25cm},clip]{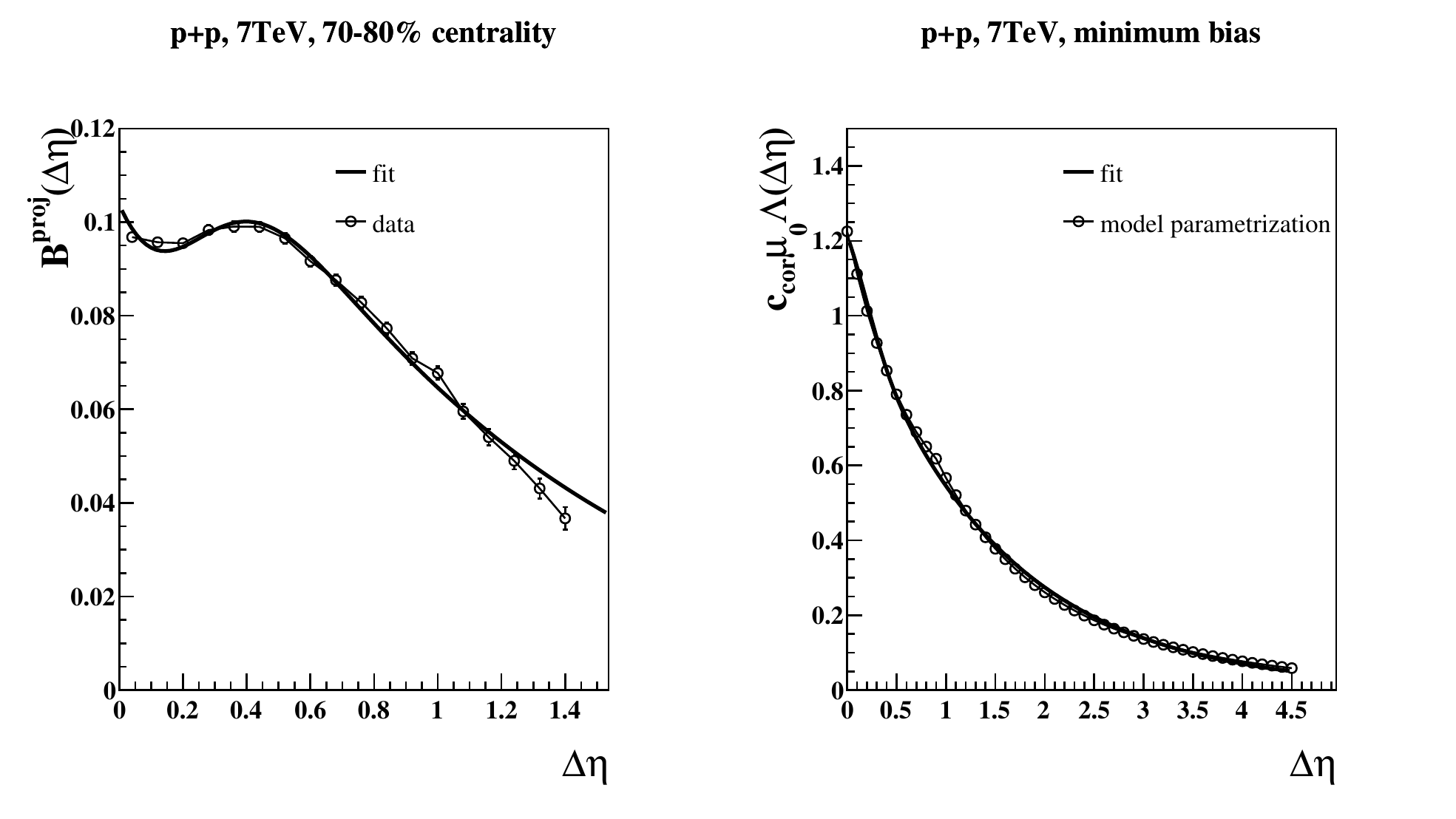}
}
\caption{\label{Balance-pp}
Left: The projection of balance function, $B^{proj}\left(\Deta\right)$,
as a function of the distance between two particles $\Deta$, measured
by the ALICE experiment \cite{ALICE16}
for 70-80$\%$ centrality class in pp collisions at 7 TeV,
together with the fit, (\ref{balance-proj}),
obtained
using the difference between
the unlike-sign and like-sign two-particle correlation functions of a string,
${\Lambda}^{+-}(\Deta)\!-\!{\Lambda}^{++}(\Deta)$.
Right: The two-particle correlation function of a string, Fig.\ref{Lam-exp},
corrected for $p_{T}$ acceptance (see text)
for all charged particles,
extracted \cite{NPA15}
from the experimental pp ALICE data on forward-backward correlations between multiplicities
at 7 TeV \cite{ALICE15},
together with the fit, (\ref{L-charge}),
obtained
using the sum of the unlike-sign and like-sign two-particle correlation functions of a string
${\Lambda}^{+-}(\Deta)\!+\!{\Lambda}^{++}(\Deta)$.
}
\end{figure}

Figure~\ref{Balance-pp} shows comparison of experimental data for 70-80$\%$ centrality class
in pp collisions at 7 TeV with the suggested fit, with parameters being listed in Table~\ref{exp-par-bal}.
\begin{table}[!tb]
 \caption[dummy]{\label{exp-par-bal}
The value of the parameters in formulae (\ref{Lam_exp_pp},\ref{Lam_exp_pm})
for the two-particle correlation functions of a string ${\Lambda}^{+-}(\Deta)$ and ${\Lambda}^{++}(\Deta)$,
obtained by a simultaneous fitting of the experimental ALICE data on balance function (BF) \cite{ALICE16} and
on forward-backward correlations (FBC) between multiplicities \cite{ALICE15} for pp collisions at 7 TeV
(see Fig. \ref{Balance-pp}).
}
  \centering
\begin{tabular}{|c|c|}
 \hline
$\sqrt{s}$,\  TeV&7.0 \\
 \hline  \hline
 $\mu_{0}\Lambda_{0}^{+-}$                    &1.42\\
 $\mu_{0}\Lambda_{0}^{++}$                    &0.76\\
 $\eta^{+-}$                    &1.34\\
 $\eta^{++}$                    &1.67\\
 $\mu_{0}\Lambda_{0}^{HBT}$                    &0.25\\
 $\eta^{HBT}$                    &0.33\\
 \hline
\end{tabular}
\end{table}

\begin{figure}[!tb]
\centering{
\includegraphics[width=115mm,,angle=0, trim={0 0.4cm 0 0.25cm},clip]{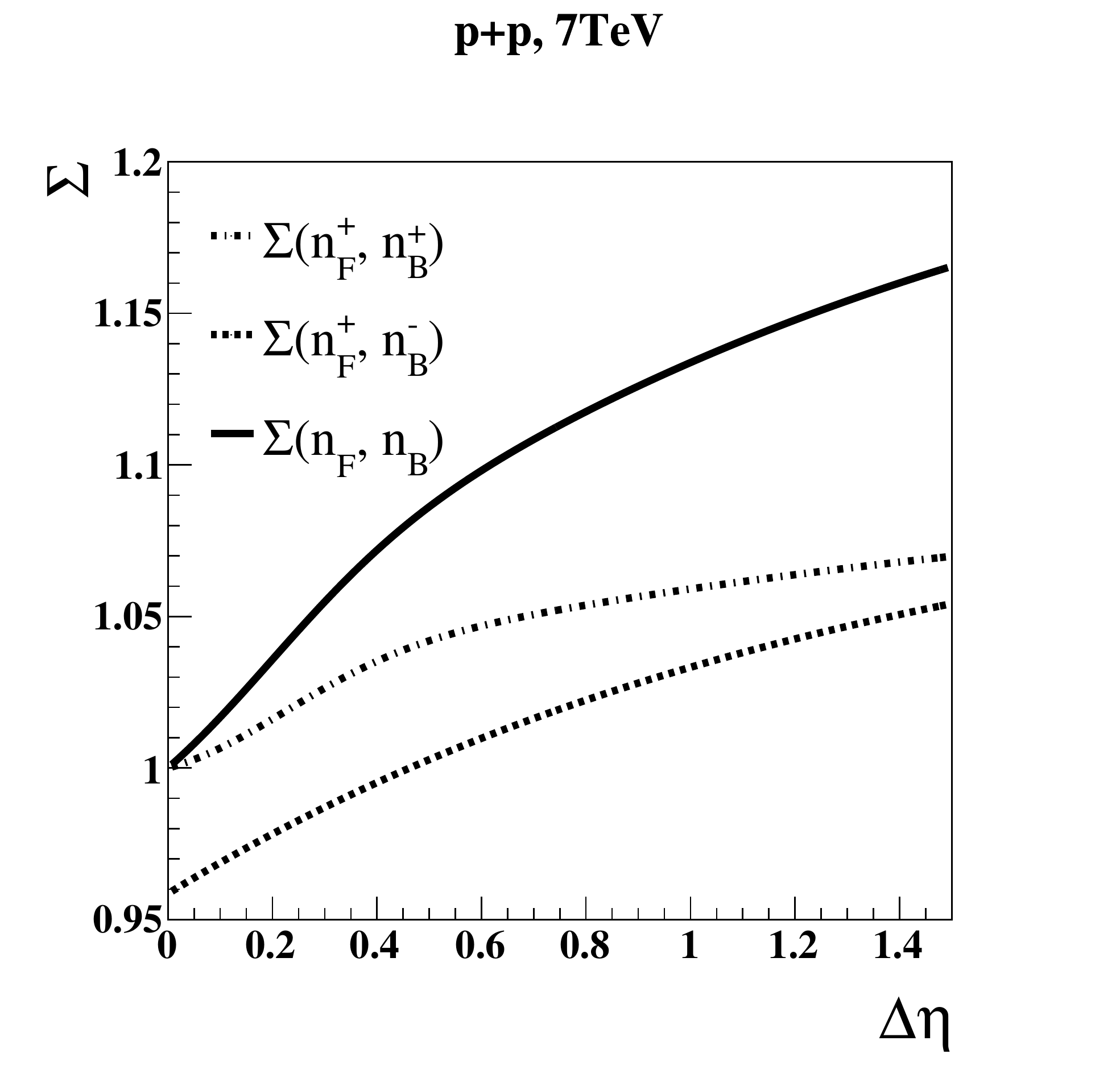}
}
\caption{\label{Sigma-pp-result}
The strongly intensive observables $\Sigma(n^{+}_{F},n^{-}_{B})$ (dashed),
$\Sigma(n^{+}_{F},n^{+}_{B})$ (dashdotted) and $\Sigma(n_{F},n_{B})$ (solid),
as a function of the distance $\Deta$ between two intervals of width $\delta\eta=0.2$,
calculated in the model with independent identical strings, using
the unlike- and like-sign two-particle correlation functions of a string, ${\Lambda}^{+-}(\Deta)$
and ${\Lambda}^{++}(\Deta)$.
}
\end{figure}

Figure~\ref{Sigma-pp-result} shows $\Sigma(n^{+}_{F},n^{-}_{B})$, $\Sigma(n^{+}_{F},n^{+}_{B})$
and $\Sigma(n_{F},n_{B})$ dependencies on $\Deta$ obtained with parameters listed in
Table~\ref{exp-par-bal}. All functions show growing behaviour with $\Deta$ with decreasing
difference between unlike-sign and like-sign strongly intensive observables.
This is a consequence of $B\left(\Deta\right)\rightarrow 0$ at large $\Deta$.
Unlike-sign $\Sigma$ is smaller than 1 at small $\Deta$,
becoming greater than 1 at larger $\Deta$. Like-sign $\Sigma$
shows behaviour that is similar to any charge sign case (see Fig.~\ref{Sigma-pp})
but suppressed in absolute value. Note that $\Sigma(n_{F},n_{B})$ was calculated here
by (\ref{Sigma-final-relation}).
It rises slightly faster than in Figure \ref{Sigma-pp} because $p_{T}$ interval was rescaled.

In Table~\ref{exp-par-bal} we see also that as one can expect from the local charge conservation
in string fragmentation process \cite{Wong15} the correlation length, $\eta^{++}$,
for the particles of same charges is larger than the one, $\eta^{+-}$, for opposite charges.

\section{$\Sigma$ with string fusion}
\label{sec-fusion}

In this section we consider the influence of processes of
interaction between strings
on the strongly intensive observable $\SigFB$.
This influence increases with initial energy
and with going from pp to heavy ion collisions.
One of the possible ways to take these processes into account
is to pass
from the model with independent identical strings
to the model with string fusion and percolation \cite{Biro84,Bialas86,BP92,BP93}.

Technically to simplify the account of the string fusion processes
one can introduce the finite lattice (the grid) in the impact parameter plane.
This approach was suggested in
\cite{Vest2} and then was successfully exploited for a description of various phenomena
(correlations, anisotropic azimuthal flows, the ridge)
in ultra relativistic nuclear collisions
\cite{EPJC04,PPR,YF07-1,YF07-2,TMF15,TMF17,BP11,BPV13,EPJA15,KV-EPJWOC14,KovYF13,KV-Dub12}.
In this approach one splits the impact parameter plane into cells, which area is equal
to the transverse area of single string and supposes the fusion of all strings
with the centers in a given cell.
This leads to
the splitting of the transverse area into domains
with different, fluctuating values of color field
within them. What is similar to the attempts to take into account
the density variation in transverse plane
in models based on the BFKL evolution \cite{LR11} and on the CGC approach \cite{KL11}.

In this model the definite set of strings of different types corresponds to the given event.
Each such string, originating from a fusion of $k$ primary strings, is characterized by its own
parameters:
the mean multiplicity per unit of rapidity, $\muk$, and the string correlation function, $\Lamk(\Deta)$.
By (\ref{SigmaJnfu}) these parameters uniquely determine the strongly intensive observable between multiplicities,
produced from decay of a string of the given type, $\Sigmuk$, defined by (\ref{SigmuFB}) and (\ref{Sigmu-s}).
For example, for two small observation windows, $\deta\ll\etakc$, separated by the rapidity distance $\Deta$,
similarly to (\ref{Sigma_small_nfu}), we have
\beq\label{Sigma_small_k}
\Sigmuk=1+\muk\deta\, [\Lamk(0)-\Lamk(\Deta)]   \ .
\eeq

In this case of the model with $k$ string types
the direct calculation
gives for the strongly intensive observable, $\SigFB$
(for symmetric reaction and observation windows $\detaF=\detaB\equiv\deta$, $\av\nF=\av\nB\equiv \av n $):
\beq\label{Sigma-w}
\SigFB
= \sum_{k=1}^{\infty}  \alpha_k\, \Sigk \ , \hs{1} \alpha_k=\frac{\av{n^{(k)}}}{\av n}   \ ,
\eeq
where
$\av{n^{(k)}}$ is a mean number of particles produced from all sells with $k$ fused strings
in the observation window $\deta$.

Note that the same result was obtained in the model with two types of strings in \cite{Andronov15}
for the long-range part of $\SigFB$,
when at $\Deta\!\gg\!\etac$ by (\ref{JFF}) and  (\ref{JFB}) we have $J_{FF}\gg J_{FB}$
and by (\ref{omn1}) and (\ref{SigmaJnfu})
$\Sigk=\omega_\mu^{(k)}$ with $k=$1,2.
What led to
\beq\label{SigmaLR}
\left.\SigFB\right |_{\Deta\gg\etac} = \frac{\avr n 1 \omega_\mu^{(1)}+\avr n 2 \omega_\mu^{(2)}}{\av n}   \ .
\eeq
One can compare this limit with the one given by the formula (\ref{limit1-2}) for the case of identical strings.

For an arbitrary rapidity distance $\Deta$
between the forward and backward observation windows of the $\deta$ width
in the model with $k$ string types we have
\beq\label{SigmaDeta}
\SigFB = 1+ \deta\sum_{k=1}^{\infty}  \alpha_k\,
\muk\, [J^{(k)}_{FF}-J^{(k)}_{FB}]  \ ,
\eeq
where we have introduced $J^{(k)}_{FF}$ and $J^{(k)}_{FB}$
for the two-partic-le correlation function $\Lamk(\eta_1-\eta_2)$
similarly to (\ref{JFF}) and (\ref{JFB}).

For narrow observation windows, $\deta\ll\etakc$, by (\ref{Sigma_small_k}),
it is simplified to
\beq\label{Sigma_smk}
\SigFB = 1+ \deta\sum_{k=1}^{\infty}  \alpha_k\,
\muk\, [\Lamk(0)-\Lamk(\Deta)]  \ .
\eeq

If we will use also the simple exponential parametrization for
$\Lamk(\Deta)$
similar to (\ref{Lam_exp}):
\beq \label{Lam_expk}
\Lamk(\Deta)=\Lamkk \exp{(-{|\Deta|}/{\etakc})}  \ ,
\eeq
then we can rewrite (\ref{Sigma_smk}) as
\beq\label{Sigma_expk}
\SigFB = 1+ \deta\sum_{k=1}^{\infty}  \alpha_k\,
\muk\Lamkk [1-\exp{(-{|\Deta|}/{\etakc})}]  \ .
\eeq
We see that in this case each string of the type $k$ is characterized by two parameters:
the product $\muk\Lamkk$, where the $\muk$ is the mean multiplicity per unit of rapidity
from a decay of such string, and
its two-particle correlation length $\etakc$,
which determines the correlations between particles,
produced from a decay of the string.

In the framework of the string fusion model \cite{Biro84,Bialas86,BP92,BP93}
one usually supposes that the mean multiplicity per unit of rapidity
for fused string, $\muk$, increase as $\sqrt{k}$ with $k$.
The dependence of the correlation length $\etakc$ on $k$ is not so obvious.
Basing on
a simple geometrical
picture of string fragmentation (see, {\it e.g.}, \cite{Artru,VENUS,Dub08,DIPSY-Tar})
one can expect the decrease of the correlation length, $\etakc$, with increase of $k$.
In this picture with a growth of string tension
the fragmentation process is finished at smaller string segments in rapidity.
The correlation takes place only between particles
originating from a fragmentation of neighbour string segments
and hence the correlation length $\etakc$ will decrease with $k$ for fused strings.

Indirectly this fact is confirmed by the analysis \cite{Titov13}
of the experimental STAR \cite{STAR-nc} and ALICE \cite{ALICE-nc} data on net-charge fluctuations in pp and AA collisions.
The dependence of net-charge fluctuations on the rapidity width of the observation window
can be well described in a string model if one supposes the decrease of the correlation length
with the transition to collisions of heavier nuclei and to higher energies, {\it i.e.} to collisions
in which the proportion of fused strings is increasing.

By (\ref{Sigma_expk}) both these factors,
the increase of $\muk$ and the decrease of $\etakc$ for fused string,
lead to the steeper increase of $\Sigk$, (\ref{Sigma_small_k}), with $\Deta$
and to its saturation at a higher level $\omega_\mu^{(k)}=1+\deta\, \muk\Lamk(0)$.
Due to (\ref{Sigma-w}) this behaviour transmits to the observable $\SigFB$,
as the last is a weighted average of $\Sigk$
with the weights
$\alpha_k={\av{n^{(k)}}}/{\av n}$, which are the mean portions of the particles
produced from a given type of strings.

In real experiment we have always a mixture of fused and single strings.
So with the transition to pp collisions at higher energy or/and to collisions of nuclei
the proportion of fused strings will increase and we will observe
the steeper increase of $\SigFB$, with $\Deta$ and its saturation at a higher level.
Really, in Fig.\ref{Sigma-pp} we see such behaviour of $\SigFB$,
when we compare $\SigFB$ for pp collisions at three initial energies: 0.9, 2.76 and 7 TeV,
obtained through fitting \cite{NPA15} of the experimental pp ALICE data \cite{ALICE15}
on forward-backward correlations between multiplicities.

Table \ref{exp-par} illustrates the increase of $\mu_0\Lambda_0$ and
the decrease of the correlation length $\etac$
with energy for this data. Note that these values are the some effective ones,
because at each energy we had supposed that all strings are identical.
So they only indirectly reflects the influence of the increase of the proportion of fused strings
with energy in pp collisions.

For studies of the $\SigFB$ dependence on multiplicity classes
we can predict the behaviour similar to the one in Fig.\ref{Sigma-pp}.
For more central pp collisions
due to the increase of the proportion of fused strings
in such collisions
we also have to observe
the steeper increase of $\SigFB$, with $\Deta$ and its saturation at a higher level.

Note that from a general point of view,
this simultaneously means
that the observable $\SigFB$,
strictly speaking, can not be considered
any more as strongly intensive.
Through the weight factors,
$\alpha_k={\av{n^{(k)}}}/{\av n}$, entering the formula
(\ref{Sigma-w}), which
are the mean portions of the particles
produced from a given type of strings,
the observable $\SigFB$
becomes dependent on collision conditions
({\it e.g.}, on the collision centrality).

\section{$\Sigma$ with PYTHIA}
\label{sec-MC}
In this section for a comparison we present
the results for the strongly intensive observables under consideration
obtained using the PYTHIA event generator.
The PYTHIA event generator \cite{Sjostrand15,Sjostrand06} with the Lund string fragmentation
model \cite{Andersson83} in its core is very successful in description of LHC data on pp collisions.
So one can expect that the results will be in correspondence with the ones obtain above
in a simple string model.

All the results to be shown below are obtained with the PYTHIA8.223 version
using its default Monash 2013 tune \cite{Skands14} from generation of 12 $\times$ $10^6$ events
for all inelastic proton-proton collisions (SoftQCD:inelastic=on). Statistical uncertainties were estimated
using the sub-sample method \cite{Efron81}, with number of sub-samples being equal to 30.
Statistical uncertainties are smaller than a line width and not visible on the plots to be presented below.
In the analysis we considered only charged particles with $0.3<p_{T}<1.5$ GeV/$c$
to be consistent with the ALICE forward-backward correlations measurements~\cite{ALICE15}.

\begin{figure}[!tb]
\centering{
\includegraphics[width=115mm,,angle=0]{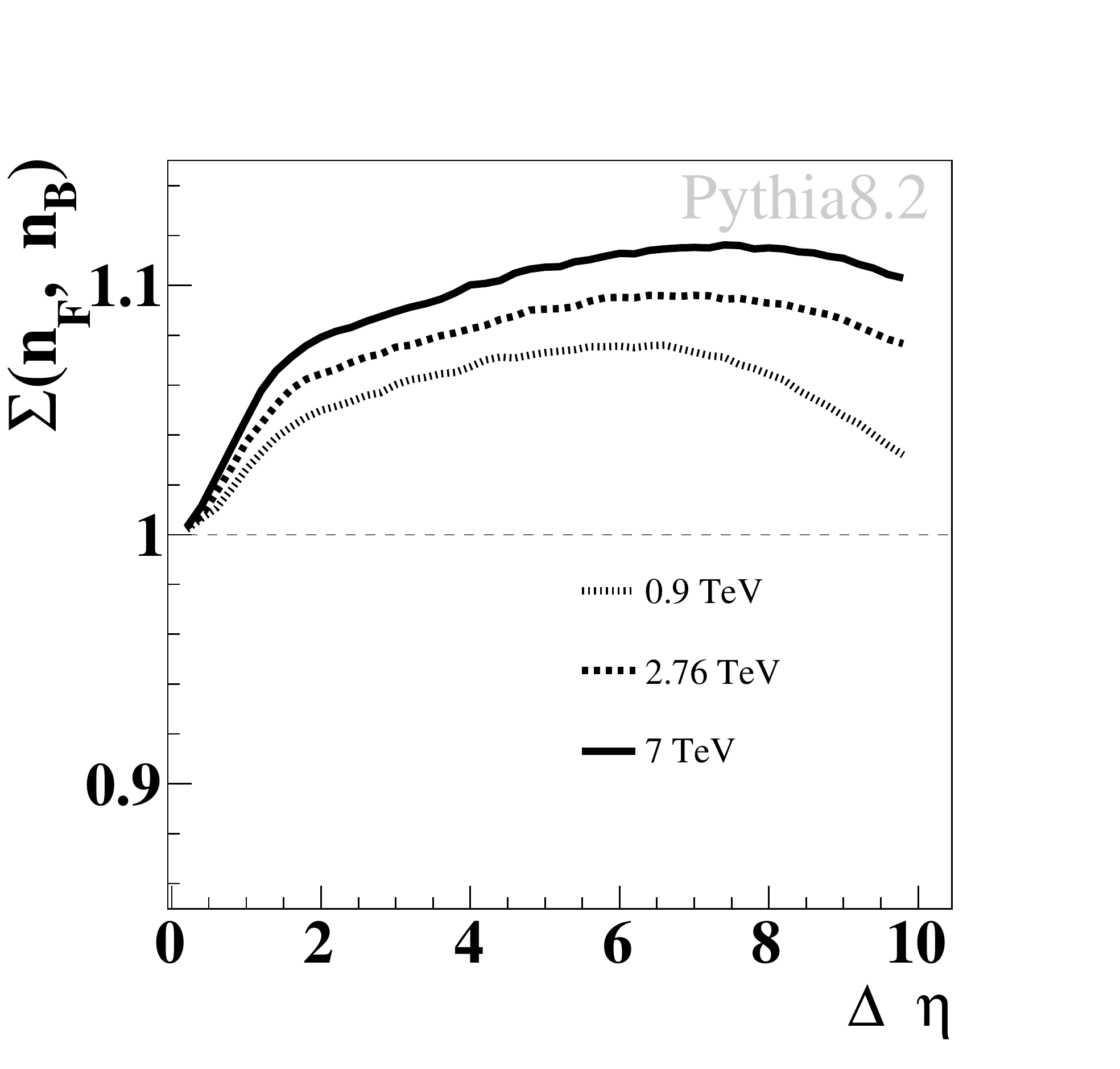}
}
\caption{\label{Sigma-pp-pythia}
The strongly intensive observable, $\SigFB$, between multiplicities of charged particles with
$0.3<p_{T}<1.5$ GeV/c
in two small pseudorapidity windows (of the width $\deta=$ 0.2)
as a function of the distance between window centers, $\Deta$, calculated
with the Monash 2013 tune of the PYTHIA8.223 model for three collision energies: 0.9, 2.76 and 7 TeV.
}
\end{figure}

Figure~\ref{Sigma-pp-pythia} shows PYTHIA8 predictions of $\SigFB$ dependencies on distance
between two windows for three collision energies.
We see that $\SigFB$ grows with separation of windows up to a certain saturation point
with a subsequent decrease. The point of saturation increases with the collision energy growth.
Moreover, for $\Delta\eta<2$ the growth is steeper than for large gaps.
The phase of growing is reminiscent of the independent string model predictions (see Fig.~\ref{Sigma-pp}).
Observed collision energy dependence is also consistent with the predictions from Sect.~\ref{sec-model}.

Change of trend at large separation between windows can be understood
as a consequence of a significant decrease of a mean multiplicities
$\langle n_F\rangle$ and $\langle n_B\rangle$ due to reduction of the number of strings
contributing to both observation windows. Such a decrease leads
to almost poissonian fluctuations of $n_F$ and $n_B$ and consequently $\SigFB\to 1$.
This effect is not taken into account in the independent string model,
where at mid-rapidities we assume that the $\langle n_F\rangle$ and $\langle n_B\rangle$
are independent of window positions, due to translation invariance in rapidity.
We supposed also that every string can contribute to both observation windows.

\begin{figure}[!tb]
\centering{
\includegraphics[width=115mm,,angle=0]{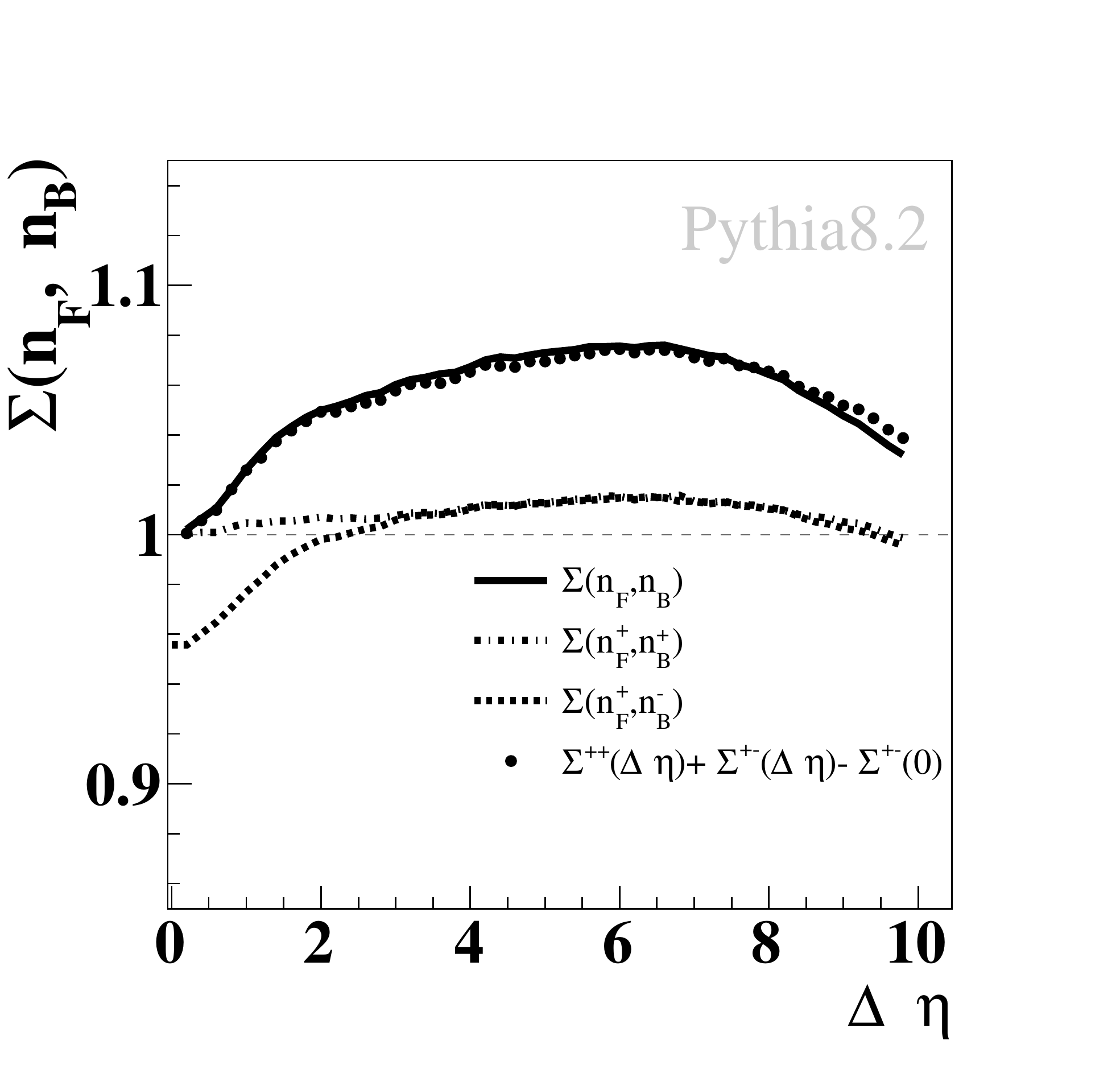}
}
\caption{\label{Sigma-pp-pythia-charge}
The strongly intensive observable, $\SigFB$, between multiplicities of charged particles with $0.3<p_{T}<1.5$ GeV/c
in two small pseudorapidity windows (of the width $\deta=$ 0.2)
as a function of the distance between window centers, $\Deta$, for 0.9 TeV collisions calculated
with the Monash 2013 tune of the PYTHIA8.223 model for different charge combinations.
For comparison the results, obtained by the formula (\ref{Sigma-final-relation}), are shown in full circles.
}
\end{figure}

Figure~\ref{Sigma-pp-pythia-charge} shows results for different combinations of electric charges.
Additional point with $\Delta\eta=0$ is calculated
for $\Sigma(n^{+}_{F},n^{-}_{B})$ in the same window ($F=B$),
see formulae (\ref{sigma-plusminus-small}) and (\ref{sigma-plusminus-null-small}).

The behavior of $\Sigma(n^{+}_{F},n^{-}_{B})$, $\Sigma(n^{+}_{F},n^{+}_{B})$
and $\Sigma(n_{F},n_{B})$ is in correspondence with the independent string model predictions
(see Figure~\ref{Sigma-pp-result}). Unlike-sign $\Sigma$, starting also from the value around 0.95 at small $\Deta$,
becomes greater than 1 at large $\Deta$. Like-sign $\Sigma$
shows behaviour that is similar to all charged case $\SigFB$, but suppressed in absolute value.

Full circles in Figure~\ref{Sigma-pp-pythia-charge} represent results obtained in PYTHIA
using the relation (\ref{Sigma-final-relation}).
Recall that this relation was obtained under the assumption of charge symmetry, (\ref{sym-ch}).
Really, one can see that this relation nicely reproduces $\SigFB$
for all charged particles (solid line) in central rapidity region,
where the charge symmetry take place at LHC energies,
while with going to a fragmentation region
it starts to fail and two curves begin to deviate.

\begin{figure}[!tb]
\centering
\includegraphics[width=70mm,,angle=0]{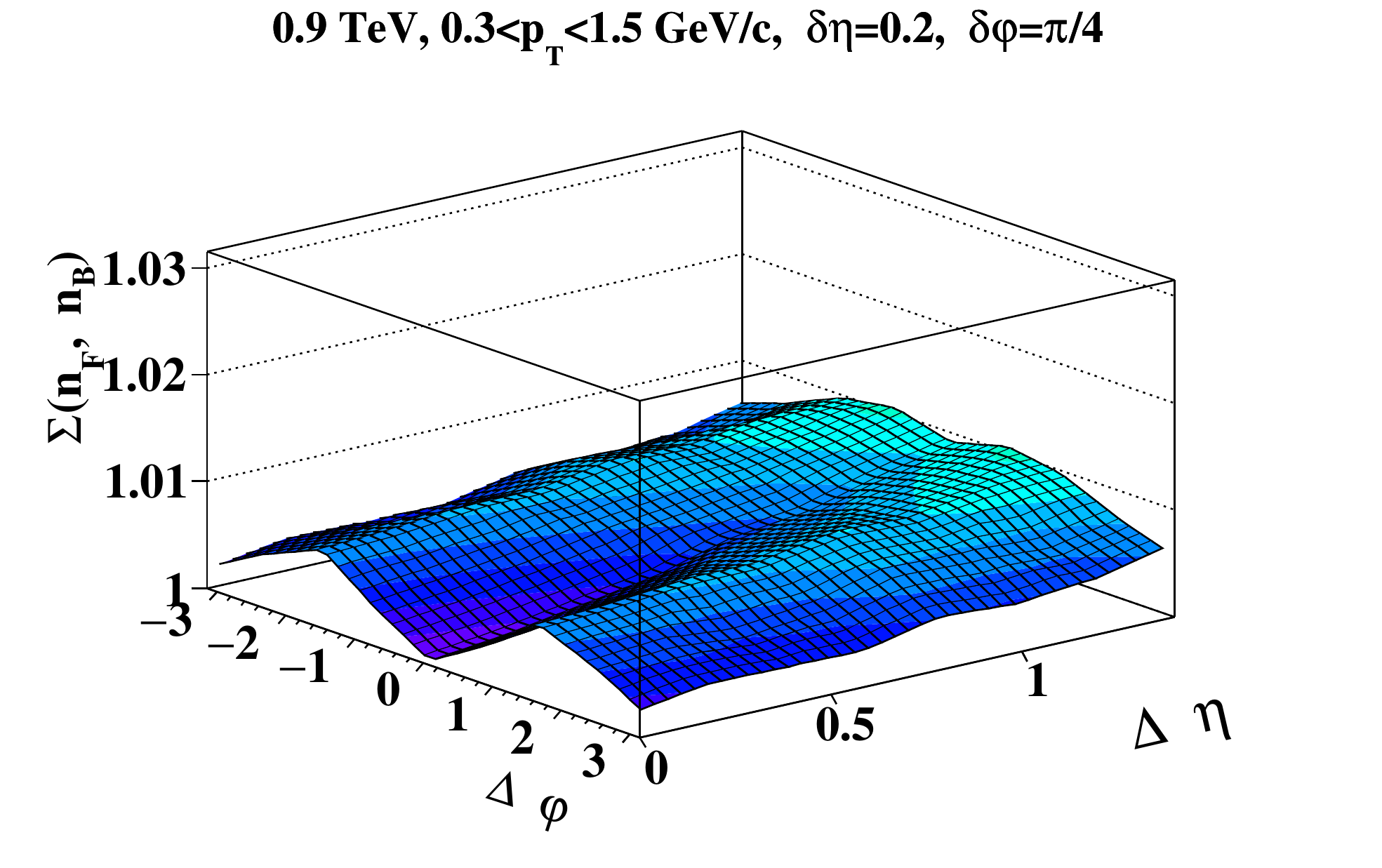}
\includegraphics[width=70mm,angle=0,clip]{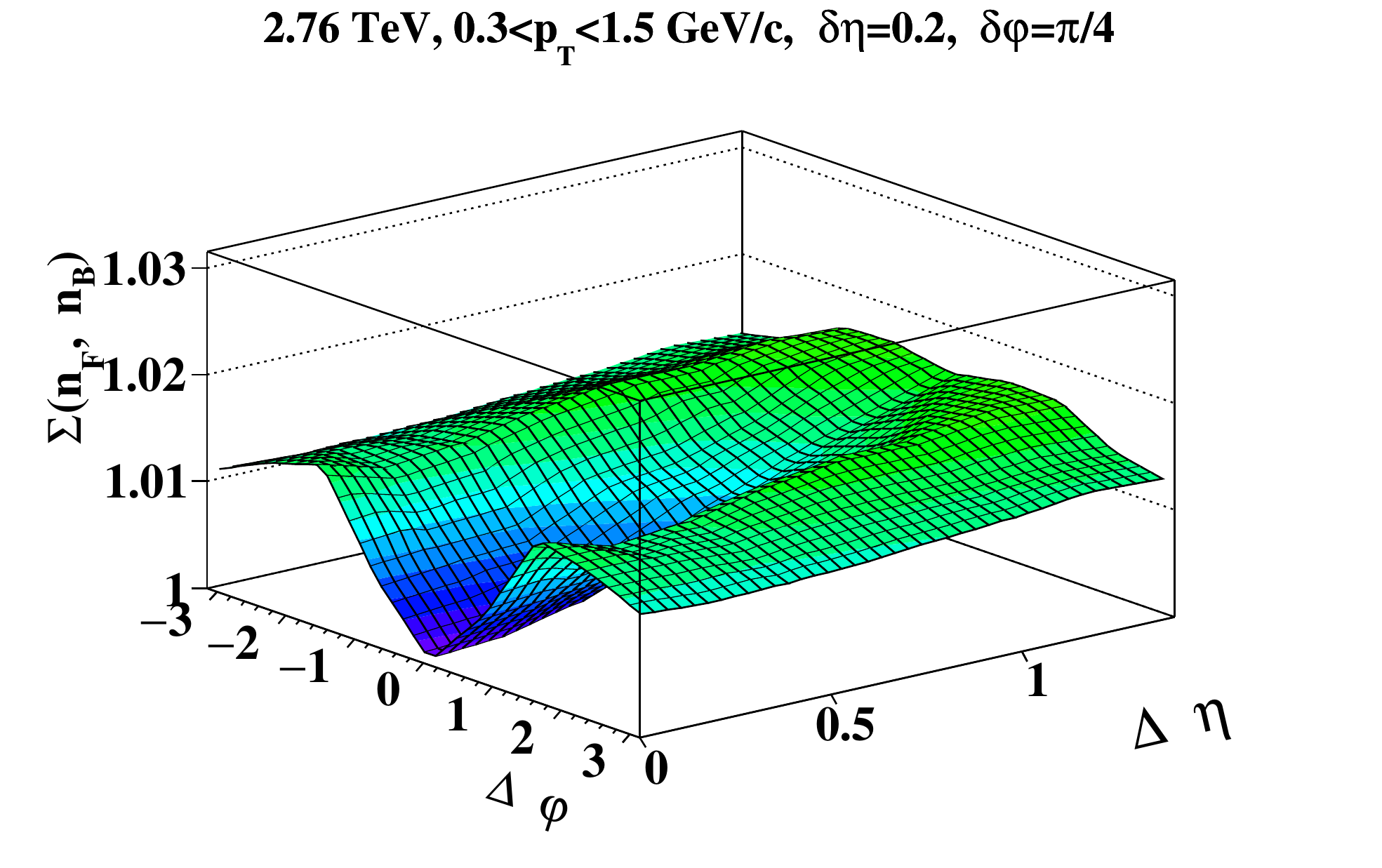}\\
\includegraphics[width=70mm,angle=0,clip]{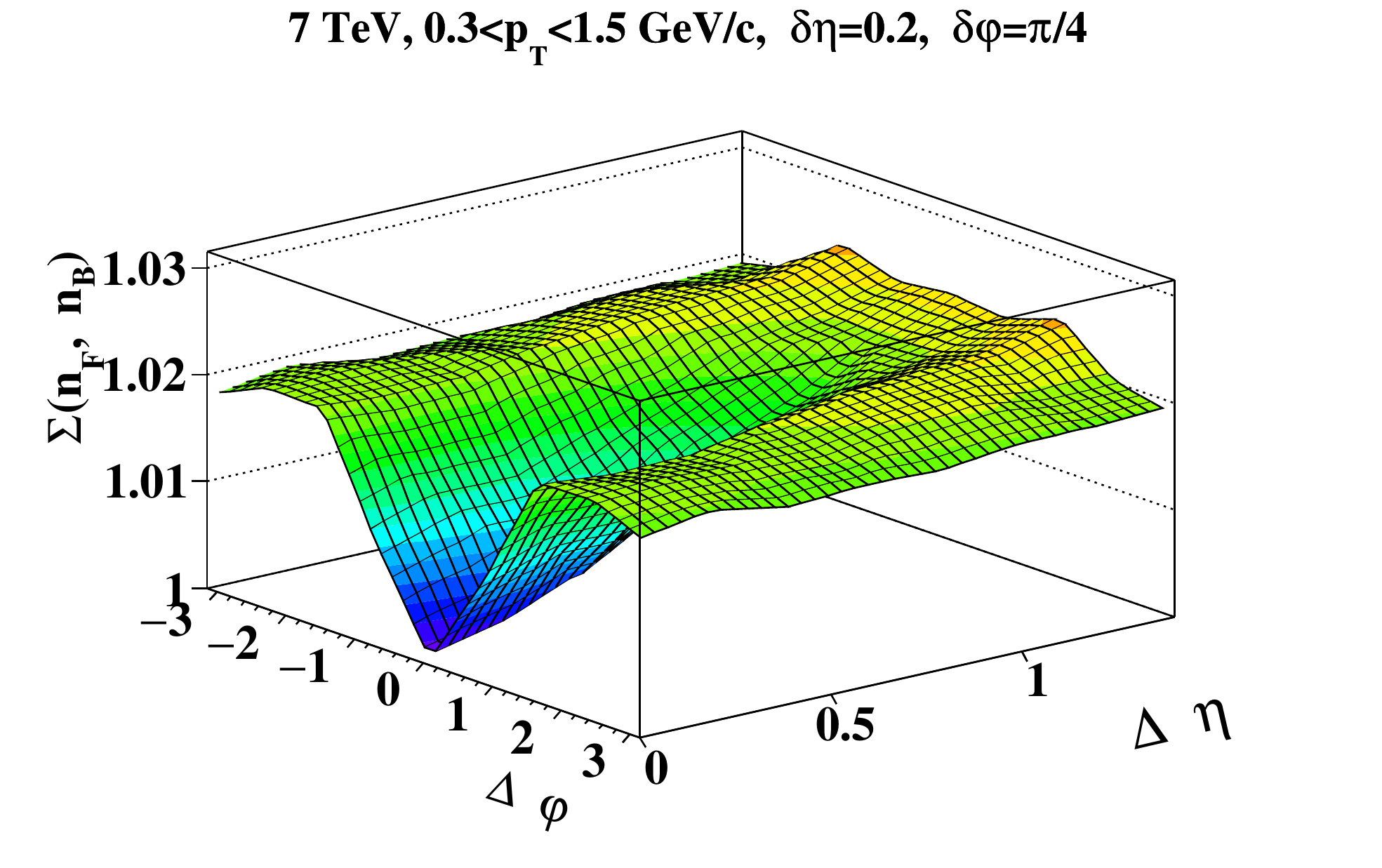}
\caption{\label{Sigma-pp-pythia-phi}
The strongly intensive observable, $\SigFB$, between multiplicities
in two small pseudorapidity - azimuthal angle windows (of the width $\deta=$ 0.2, $\delta\varphi=\pi/4$)
as a function of the distance between window centers, $\Deta$, $\Delta\varphi$, calculated
with the Monash 2013 tune of the PYTHIA8.223 model for three collision energies: 0.9, 2.76 and 7 TeV.
}
\end{figure}

Results obtained with the PYTHIA8 event generator for $\SigFB$ for windows separated both
in pseudorapidity and azimuthal angle are presented in Fig.~\ref{Sigma-pp-pythia-phi}.
The shape of obtained functions - a dip at small values of $\Delta\eta$ and $\Delta\phi$ and
a plateau at larger values - is again in qualitative
agreement with the independent string model
predictions (see Fig.~\ref{Sigma-dd1}).
It is important to note
that in the framework of the simple string model,
in contrast with PYTHIA calculations,
we can clearly see
the physical reasons for such behavior.

\section{Summary and conclusions}
\label{Concl}

The using of strongly intensive observables are considered
as a way to suppress the contribution of trivial "volume" fluctuations
in experimental studies of the correlation and fluctuation phenomena  (see, {\it e.g.} \cite{GorGaz11}).
In present paper we have studied
the properties of strongly intensive observable between multiplicities
in two acceptance windows separated in rapidity and azimuth, $\SigFB$,
in the model with quark-gluon strings (color flux tubes) as sources.

We show that in the case with independent identical strings
the strongly intensive character of this observable is being confirmed:
it depends only on the individual characteristics of a string and
is independent of both
the mean number of strings and its fluctuation.
These individual characteristics of a string are
a mean number of particles per unit of rapidity, $\mu_0$,
produced from string fragmentation,
and the two-particle correlation function, $\Lambda(\Deta,\Dphi )$,
characterizing the correlations between particles, produced from
the same string.

The ALICE experimental data \cite{ALICE15} on forward-back-ward correlations (FBC)
in small rapidity windows separated in rapidity and azimuth enables to obtain
the information on this string correlation function, $\Lambda(\Deta,\Dphi )$, \cite{NPA15}.
Using it we calculate the dependence of the strongly intensive observable, $\SigFB$, 
on the acceptance of observation windows and the gaps between them.

We have studied also
the strongly intensive observables between multiplicities
taking into account the sigh of particle charge,
$\Sigma(\nFp, \nBp)$, $\Sigma(\nFp, \nBm)$ and $\Sigma(\nFp, \nFm)$.
We express them through the string correlation functions
between like and unlike charged particles, $\Lambda^{++}(\Deta)$ and ${\Lambda}^{+-}(\Deta)$.
To calculate these quantities we need more information
on a string decay process, because the FBC data \cite{ALICE15} contains
information only on the sum of these correlation functions,
$\Lambda(\Deta)=[\Lambda^{++}(\Deta)+{\Lambda}^{+-}(\Deta)]/2$.

We show that the so-called balance function (BF)
can be expressed through the difference of these string correlation functions.
Using the ALICE experimental data \cite{ALICE16} on BF we
extract both correlation functions between like and unlike charged particles
produced from a fragmentation of a single string.
With these correlation functions we calculate the properties of
the strongly intensive observables,
$\Sigma(\nFp, \nBp)$, $\Sigma(\nFp, \nBm)$ and $\Sigma(\nFp, \nFm)$.
In particular we found that as one can expect from the local charge conservation
in string fragmentation process \cite{Wong15} the correlation length
for the particles of same charges is larger than the one for opposite charges.

In the case when the string fusion processes
are taken into account and a formation
of strings of a few different types takes place in a collision
we show that the observable $\SigFB$
is proved to be equal to a weighted
average of its values for different string types.
Unfortunately in this case
through the weight factors
this observable becomes dependent on collision conditions
and, strictly speaking, can not be considered
any more as strongly intensive variable.
This complicates the quantitative predictions for the $\SigFB$,
as it starts to depend on experimental conditions through
these weight factors.

Nevertheless we argue that the string fusion leads to the
following changes of individual string characteristics:
the multiplicity density per unit of rapidity
from fragmentation of fused string occurs higher
and the correlation length between particles, produced from
the fused string, becomes smaller.
Both these factors lead to the steeper increase of $\SigFB$
with rapidity gap between windows, $\Deta$,
and to its saturation at a higher level,
what is qualitatively consistent with available experimental data
on pp collisions at different LHC energies.

We compare our results also with the PYTHIA8 event generator predictions.
We find a similar picture for the dependence
of $\Sigma(\nFp, \nBp)$, $\Sigma(\nFp, \nBm)$ and $\Sigma(\nF, \nB)$
on rapidity gap between windows, $\Deta$, at mid-rapidities,
where the exploited boost invariant version of string model is applicable.

\section*{Acknowledgements}
\label{Ackn}

The research was funded by the grant of the Russian Science Foundation (project 16-12-10176).
%

\begin{thebibliography}{99}
\bibitem{Dumitru08}
A. Dumitru, F. Gelis, L. McLerran, R. Venugopalan,
Nucl. Phys. A \textbf{810}, (2008) 91.
\bibitem{Biro84} T.S. Biro, H.B. Nielsen, J. Knoll, {Nucl. Phys. B} \textbf{245}, (1984) 449.
\bibitem{Bialas86} A. Bialas, W. Czyz, {Nucl. Phys. B} \textbf{267}, (1986) 242.
\bibitem{BP92}
M.A. Braun, C. Pajares, Phys. Lett. B \textbf{287}, (1992) 154.
\bibitem{BP93}
M.A. Braun, C. Pajares, Nucl. Phys. B \textbf{390}, (1993) 542.
\bibitem{PRL94}
N.S. Amelin {\it et al.},
Phys. Rev. Lett. \textbf{73}, (1994) 2813.
\bibitem{BP00} M.A. Braun, C. Pajares,
{Phys. Rev. Lett.} \textbf{85}, (2000) 4864.
\bibitem{EPJC04}
M.A. Braun, R.S. Kolevatov, C. Pajares, V.V. Vechernin,
Eur. Phys. J. C \textbf{32}, (2004) 535.
\bibitem{PPR}
ALICE collaboration {\it et al.},
J. Phys. G \textbf{32}, 10, (2006) 1295.
\bibitem{YF07-1}
V.V. Vechernin, R.S. Kolevatov, Phys. Atom. Nucl. \textbf{70}, (2007) 1797.
\bibitem{YF07-2}
V.V. Vechernin, R.S. Kolevatov, Phys. Atom. Nucl. \textbf{70}, (2007) 1809.
\bibitem{TMF15}
V.V. Vechernin,
Theor. Math. Phys., \textbf{184}, (2015) 1271.
\bibitem{TMF17}
V.V. Vechernin,
Theor. Math. Phys., \textbf{190}, (2017) 251.

\bibitem{GorGaz11}
M.I. Gorenstein, M. Gazdzicki, Phys. Rev. C, \textbf{84}, (2011) 014904.

\bibitem{PLB00}
M.A. Braun, C. Pajares, V.V. Vechernin,
Phys. Lett. B \textbf{493}, (2000) 54.
\bibitem{CapKrz78}
A. Capella, A. Krzywicki, Phys. Rev. D \textbf{18}, (1978) 4120.
\bibitem{NPA15} V.~Vechernin, Nucl. Phys. A \textbf{939}, (2015) 21.
\bibitem{Voloshin02}
C. Pruneau, S. Gavin, and S. Voloshin,
\emph{Phys. Rev. C} {\bfseries 66} (2002) 044904.

\bibitem{Uhlig78}
S. Uhlig, I. Derado, R. Meinke, H. Preissner,
\emph{Nucl. Phys. B} \textbf{132} (1978) 15.

\bibitem{Derrick86}
M. Derrick {\it et al.},
\emph{Phys. Rev. D} \textbf{34} (1986) 3304.

\bibitem{Back06}
B.B. Back {\it et al.} (PHOBOS Collaboration),
\emph{Phys. Rev. C} \textbf{74} (2006) 011901.

\bibitem{ALICE15} J. Adam {\it et al.} (ALICE Collaboration),
JHEP \textbf{05}, (2015) 097.

\bibitem{ALICE16} J. Adam {\it et al.} (ALICE Collaboration),
Eur. Phys. J. C \textbf{76}, (2016) 86.

\bibitem{Sjostrand15}
T. Sjostrand {\it et al.}, Comput. Phys.Commun. \textbf{191}, (2015) 159.
\bibitem{Sjostrand06}
T. Sjostrand, S. Mrenna and P. Skands, \emph{JHEP} \textbf{05}, (2006) 026.

\bibitem{Andronov16}
E. Andronov (for the NA61/SHINE Collaboration)
\emph{J. Phys. Conf. Ser.} {\bfseries 668}, (2016)
012036.

\bibitem{Wong15}
C.-Y. Wong, \emph{Phys. Rev. D} {\bfseries 92}, (2015) 074007.


\bibitem{Vest2} V.V. Vechernin, R.S. Kolevatov,
Vestn. Peterb. Univ., Ser.4: Fiz. Khim. \textbf{4}, (2004) 11, arXiv: hep-ph/0305136.
\bibitem{BP11} M.A. Braun, C. Pajares, Eur. Phys. J. C \textbf{71}, (2011) 1558.
\bibitem{BPV13} M.A. Braun, C. Pajares, V.V. Vechernin, Nucl. Phys. A \textbf{906}, (2013) 14.
\bibitem{EPJA15} M.A. Braun, C. Pajares, V.V. Vechernin, Eur. Phys. J. A \textbf{51}, (2015) 44.


\bibitem{KV-EPJWOC14}
V. Kovalenko, V. Vechernin,
EPJ Web of Conferences \textbf{66}, (2014) 04015.
\bibitem{KovYF13}
V.N. Kovalenko, Phys. Atom. Nucl. \textbf{76}, (2013) 1189.
\bibitem{KV-Dub12}
V. Kovalenko, V. Vechernin,
Proceedings of Science \textbf{Baldin-ISHEPP-XXI}, (2012) 077.


\bibitem{LR11}
E. Levin, A.H. Rezaeian, Phys.Rev. D \textbf{84}, (2011) 034031.
\bibitem{KL11}
A. Kovner, M. Lublinsky, Phys.Rev. D \textbf{83}, (2011) 034017.

\bibitem{Andronov15}
E.V. Andronov,
Theor. Math. Phys., \textbf{185},  (2015) 1383.

\bibitem{Artru} X. Artru, Phys. Rept. {\bf 97}, (1983) 147.
\bibitem{VENUS}
K. Werner,
Phys. Rept. {\bf 232}, (1993) 87.

\bibitem{Dub08} V.V. Vechernin,
Proceedings of the Baldin ISHEPP XIX vol.1, JINR, Dubna (2008) 276-281; arXiv:0812.0604.

\bibitem{DIPSY-Tar}
C. Bierlich, G. Gustafson, L. Lonnblad, A. Tarasov,
JHEP \textbf{03}, (2015) 148,
arXiv:1412.6259.


\bibitem{Titov13} A.~Titov, V.~Vechernin,
Proceedings of Science \textbf{Baldin-ISHEPP-XXI}, (2013) 047.
\bibitem{STAR-nc} B.I. Abelev {\it et al.} (STAR Collaboration),
Phys. Rev. C \textbf{79}, (2009) 024906.
\bibitem{ALICE-nc} B. Abelev {\it et al.} (ALICE Collaboration),
Phys. Rev. Lett. \textbf{110}, (2013) 152301.
\bibitem{Andersson83}
B. Andersson {\it et al.}, Phys. Rept. \textbf{97}, (1983) 31.
\bibitem{Skands14}
P. Skands, S. Carrazza, J. Rojo, Eur.Phys.J. C \textbf{74}, 8 (2014) 3024.
\bibitem{Efron81}
B. Efron, Biometrika \textbf{68}, 3 (1981) 589.
\end{thebibliography}
%

\end{document}